\documentclass[amssymb,aps, prl,superscriptaddress, 10pt]{revtex4-2}

\usepackage{amsmath}
\usepackage{fullpage}
\usepackage{amssymb}
\usepackage{amsfonts}
\usepackage{color}
\usepackage[pdftex]{graphicx}
\usepackage{epstopdf}
\usepackage{soul}
\usepackage{xcolor}
\usepackage{multirow} 
\usepackage{caption}
\usepackage{subcaption}
\usepackage{cases}
\usepackage{url}
\newcommand{\hfds}{HfS$_2$}
\newcommand{\xhat}{\widehat{\mathbf{x}}}
\begin{document}

\title{Computational discovery of high-refractive-index van der Waals materials: The case of \hfds} 

\author{Xavier Zambrana-Puyalto}
\email{xavislow@protonmail.ch}
\affiliation{Department of Physics, Technical University of Denmark, Fysikvej, DK-2800 Kongens Lyngby, Denmark}
\author{Mark Kamper Svendsen} 
\affiliation{NNF Quantum Computing Programme, Niels Bohr Institute, University of Copenhagen, Denmark}
\author{Amalie H. Søndersted}
\affiliation{Department of Physics, Technical University of Denmark, Fysikvej, DK-2800 Kongens Lyngby, Denmark}
\author{Avishek Sarbajna}
\affiliation{Department of Physics, Technical University of Denmark, Fysikvej, DK-2800 Kongens Lyngby, Denmark}
\author{Joakim P. Sandberg}
\affiliation{Department of Physics, Technical University of Denmark, Fysikvej, DK-2800 Kongens Lyngby, Denmark}
\author{Albert L. Riber}
\affiliation{Department of Physics, Technical University of Denmark, Fysikvej, DK-2800 Kongens Lyngby, Denmark}
\author{Georgy Ermolaev}
\affiliation{Emerging Technologies Research Center, XPANCEO, Internet City, Emmay Tower, Dubai, United Arab Emirates}
\author{Tara Maria Boland}
\affiliation{Department of Physics, Technical University of Denmark, Fysikvej, DK-2800 Kongens Lyngby, Denmark}
\author{Gleb Tselikov}
\affiliation{Emerging Technologies Research Center, XPANCEO, Internet City, Emmay Tower, Dubai, United Arab Emirates}
\author{Valentyn S. Volkov}
\affiliation{Emerging Technologies Research Center, XPANCEO, Internet City, Emmay Tower, Dubai, United Arab Emirates}
\author{Kristian S. Thygesen}
\affiliation{Department of Physics, Technical University of Denmark, Fysikvej, DK-2800 Kongens Lyngby, Denmark}
\author{Søren Raza}
\email{sraz@dtu.dk}
\affiliation{Department of Physics, Technical University of Denmark, Fysikvej, DK-2800 Kongens Lyngby, Denmark}

\begin{abstract}
New high-refractive-index dielectric materials may enhance many optical technologies by enabling efficient manipulation of light in waveguides, metasurfaces, and nanoscale resonators. Van der Waals materials are particularly promising due to their excitonic response and strong in-plane polarizability. Here we combine ab initio calculations and experiments to discover new high-refractive-index materials. Our screening highlights both known and new promising optical materials, including hafnium disulfide (HfS$_2$), which shows an in-plane refractive index above 3 and large anisotropy in the visible range. We confirm these theoretical predictions through ellipsometry measurements and investigate the photonic potential of HfS$_2$ by fabricating nanodisk resonators, observing optical Mie resonances in the visible spectrum. Over the course of seven days, we observe a structural change in HfS$_2$, which we show can be mitigated by storage in either argon-rich or humidity-reduced environments. This work provides a comparative overview of high-index van der Waals materials and showcases the potential of HfS$_2$ for photonic applications in the visible spectrum.
\end{abstract}

\maketitle

\section{Introduction}
The complex refractive index is a fundamental material property in dielectric nanophotonics that directly influences the performance of optical devices like metasurfaces and optical antennas. A high refractive index with minimal optical absorption is crucial for optimizing these technologies. For instance, metalenses achieve better focusing efficiency and achromatic metasurfaces broaden their operational bandwidth with increasing refractive index~\cite{Bayati2019,Yang2017}. Similarly, the quality factor of Mie resonances and supercavity modes in optical resonators scales strongly with the refractive index~\cite{Rybin2017,Zambrana2024}. Moreover, Miller's rule links a higher refractive index to a stronger second-order nonlinear response, highlighting its importance in nonlinear optics~\cite{BoydBook}.

Despite the critical role of the refractive index, current research predominantly relies on materials like silicon and III-V semiconductors~\cite{Baranov2017,Yang2023}. These materials offer low-loss operation but are limited to the infrared and red parts of the visible spectrum, constrained by their band gap energies. Expanding the operational range to the entire visible spectrum requires materials with a larger band gap energy to avoid optical losses, but this typically results in a lower refractive index. This effect is captured by the so-called Moss rule~\cite{Moss1950,Moss1985}, which shows that there is an inverse relation between the refractive index and the band gap energy. Consequently, there is a shortage of high-refractive-index materials that can operate effectively across the visible spectrum. Identifying and utilizing new transparent dielectric materials with high refractive indices across the visible spectrum could significantly enhance the performance and functionality of nanoscale optical devices~\cite{Khurgin2022}.

Bulk van der Waals (vdW) materials are promising in this regard, as they display weak interlayer bonding and strong covalent in-plane bonding as well as host prominent in-plane excitonic transitions \cite{Ermolaev2021}. These key properties result in an increased in-plane refractive index and suggest that vdW materials with large band gap energies may still have a high refractive index~\cite{Khurgin2022}. Recent measurements have indeed shown that vdW materials, such as transition metal dichalcogenides (TMDCs), possess very high in-plane refractive indices in the red and in the near-infrared spectral ranges, surpassing their isotropic counterparts, such as silicon~\cite{Verre2019,Munkhbat2022}. With only few experimentally characterized~\cite{Munkhbat2022,Zotev2023} and more than 1000 theoretically predicted~\cite{Mounet2018,Haastrup2018}, vdW materials could provide the needed material platform to engineer light across the entire visible spectrum. 

The search for novel optical materials can be dramatically accelerated by employing first principles electronic structure calculations based on density functional theory (DFT). With such methods, it is possible to scrutinize the properties of crystalline materials much more quickly than by traditional experiments, as successfully demonstrated in the areas of catalysis~\cite{Nørskov2009}, batteries~\cite{Xiao2019}, and solar cells~\cite{Yu2012,Kuhar2017}. However, apart from a few studies~\cite{Naccarato2019,Svendsen2022}, the high-throughput strategy has not been used to search for improved optical materials. The reason for this is that optical properties, such as the refractive index tensor, involve excited electronic states that are orders of magnitude more demanding in terms of computational resources. The current gold standard for first-principles calculation of optical properties is the GW-BSE method~\cite{Albrecht1998}, which accounts for excitonic effects. GW-BSE accurately predicts absorption spectra, but it can lead to a significant underestimation of the refractive index. To improve this, some of us have recently developed the BSE+ method, which goes beyond the BSE method and significantly improves the agreement with experiments for the refractive index~\cite{Søndersted2024}. These advancements enable us to identify new optical materials with high fidelity.

\begin{figure}
    \centering
    \includegraphics[width=0.9\textwidth]{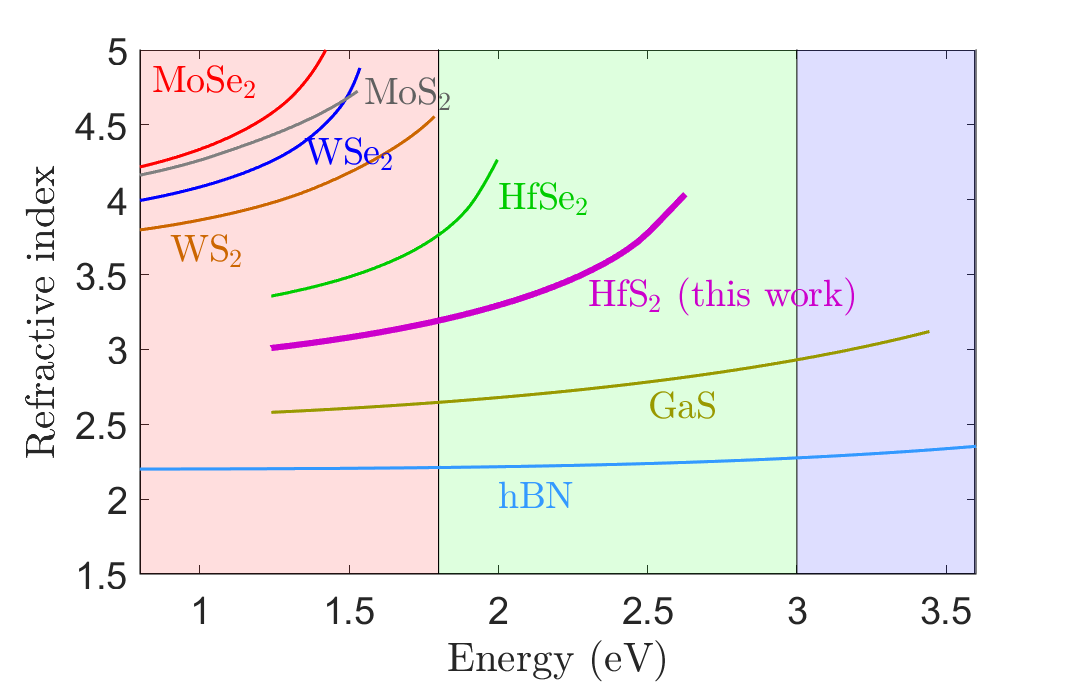}
    \caption{\textbf{vdW materials in optics.} In-plane refractive index for the most commonly used vdW materials in optics along with the new TMDC \hfds\ discovered in this work. Refractive index data are obtained from Ref.~\cite{Lee2019, Munkhbat2022, Vyshnevyy2023, Zotev2023, Polyanskiy2024} and plotted in the energy  range where the imaginary part of the refractive index, i.e. the extinction coefficient, is below 0.1.}
    \label{fig:TMDs}
\end{figure}

In this work, we use high-throughput DFT calculations to screen the optical properties of 338 semiconductor materials, starting from an initial set of 1693 unary and binary materials. Of these semiconductors, 131 materials have anisotropic refractive indices, many of them being van der Waals materials in nature. The screening identifies many super-Mossian materials~\cite{Doiron2022} that surpass the Moss rule, suggesting enhanced refractive index compared to state-of-the-art materials. In particular, the TMDC hafnium disulfide (HfS$_2$) stands out due to its low optical losses and its greater than 3 refractive index in the visible spectral range (Fig.~\ref{fig:TMDs}). To the best of our knowledge, the optical properties of this material for photonics have not been previously studied. We therefore conduct a thorough experimental optical study of \hfds\ focusing on both its bulk properties and its behavior when patterned as nanoscale Mie resonators. Using imaging ellipsometry, we measure both the in-plane and out-of-plane complex refractive indices of \hfds, confirming the BSE+ calculations of low losses and high refractive index. We also exfoliate \hfds\ and develop a fabrication procedure to realize \hfds\ nanodisks that support optical Mie resonances. We observe that \hfds\ is chemically unstable at ambient conditions, but this instability can be circumvented by storing the material in oxygen-free or humidity-reduced environments. Overall, our results demonstrate that \hfds\ has a strong potential for photonic applications due to its high refractive index and low absorption. In addition, our DFT screening provides a comparative overview of promising high-index materials that could be taken further for experimentation.

\section{Results}
\subsection{High-throughput screening and refractive index measurement}
Our high-throughput screening is based on the procedure developed in our previous work~\cite{Svendsen2022}. Briefly, we extract 1693 elementary and binary materials from the Open Quantum Materials Database~\cite{kirklin2015open} and relax the atomic structure using DFT with the Perdew--Burke--Ernzerhof (PBE) exchange correlation functional~\cite{PBE} and the D3 correction to account for the van der Waals forces~\cite{DFTD3}. Electronic band gaps for all of the materials are calculated and materials identified as metals are discarded. This leaves 338 semiconductors for which we calculate the refractive index tensor within the random phase approximation (RPA). The refractive index tensor of all of these materials are available in the newly established CRYSP database \cite{CRYSPdatabase}. We compute the fractional anisotropy to categorize the materials based on the anisotropy of their refractive index tensor. Isotropic materials are excluded, leaving 131 anisotropic materials, many of which are van der Waals in nature (see Methods for more details).

\begin{figure}
    \centering
    \includegraphics[width=1\textwidth]{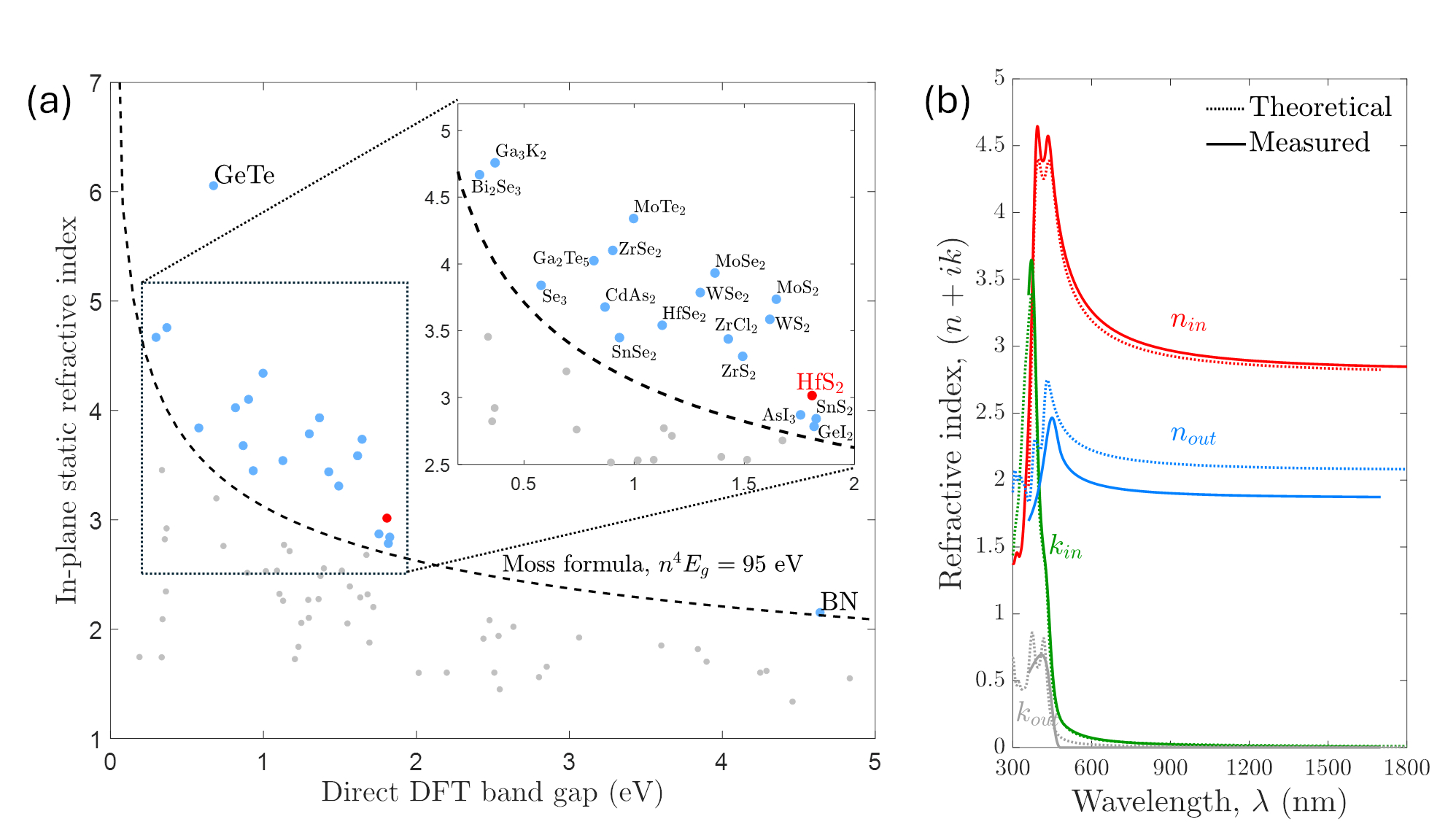}
    \caption{\textbf{Identification of high-refractive-index vdW materials.} (a) Static in-plane refractive index as a function of direct band gap energy for 72 anisotropic semiconductor materials along with the Moss formula. Chemical formulas are provided for all of the super-Mossian materials. (b) Computed (dashed lines) and experimentally measured (solid lines) in-plane (red) and out-of-plane (blue) refractive indices. The in-plane and out-of-plane extinction coefficients are also plotted in green and grey, respectively.}
    \label{fig:RI_BG}
\end{figure}

The static in-plane refractive index as a function of the direct band gap energy of 72 of all 131 materials are presented in Fig.~\ref{fig:RI_BG}(a). We have chosen the 72 materials whose two different in-plane refractive index components differ by less than 1\%. The remaining 59 materials, which display anisotropy in all three directions, are presented in the Supplementary Note~1. We observe that the data points qualitatively follow the Moss relation, given by $n^4E_g=95$~eV~\cite{Moss1950}. However, there are many materials that surpass the Moss rule, the so-called super-Mossian materials~\cite{Doiron2022}, which we label by their chemical formulas in Fig.~\ref{fig:RI_BG}(a). When examining this map of materials, it is worth noting that the PBE functional used in this work is known to systematically underestimate the band gap energy for certain materials by up to 1 eV or more~\cite{perdew1985density}. This can be remedied by employing more accurate and computationally expensive many-body calculation methods, which we present later. Nonetheless, this screening verifies that known high-index vdW materials, such as WS$_2$ and MoS$_2$, are indeed super-Mossian, and also showcases a wide range of promising but understudied optical materials, such as SnS$_2$, ZrS$_2$, and HfS$_2$. In addition, we note that for photonic applications at telecom energies ($0.8-0.95$~eV), there are many candidate materials which significantly surpass traditional semiconductor materials and could enable enhanced performance and miniaturization in, e.g., integrated photonic circuits. In this work, we direct attention to the visible spectrum, where the lack of super-Mossian materials is particularly acute. We therefore focus on the super-Mossian material with the simultaneously highest refractive index and widest band gap energy, which, based on this screening, is hafnium disulfide (\hfds). 

%The results of our simulations can be seen in Figure \ref{fig:RI_BG}(a), where the Moss rule is also plotted. We observe that almost all the super-Mossian materials exist for $E_g < 2 \mathrm{eV} \approx 620$ nm. For optical applications, we are interested in finding materials with low losses in the visible, that is why we focus on one of the super-Mossian materials with the greatest $E_g$ and RI, i.e. hafnium disulfide - whose crystal structure is depicted in Figure \ref{fig:RI_BG}(b) for completeness. Figure \ref{fig:RI_BG}(a) does not give any spectral information, it only gives us a rough estimation of the static RI of \hfds. Besides, it is known that the method used to calculate the static RI consistently yields RI values that are below/above(\textcolor{red}{Is it above or below?}) the real RI (see Methods). 

To assess the potential of \hfds\ with high fidelity, we perform a many-body BSE+ calculation of its refractive index tensor (see Methods). The BSE+ in-plane and out-of-plane complex refractive indices of \hfds\ are shown in Fig.~\ref{fig:RI_BG}(b) (dashed lines). We observe that the extinction coefficient $k$ of both in-plane and out-of-plane components have values below 0.1 for wavelengths greater than 550~nm. We also see that the theoretical in-plane refractive index $n_\textrm{in}$ is above 3 across the whole visible range, even reaching values of 4.5 in the blue spectral range. The out-of-plane refractive index $n_\textrm{out}$ varies between 2.1 and 2.7 in the visible range, highlighting a strong anisotropy of $n_\textrm{in}-n_\textrm{out} \gtrsim 1 $. These theoretical results show that the combination of high-throughput screening combined with many-body BSE+ calculations provide a rational approach to discover new optical materials with high fidelity. 

%Thus, we implement a second refined DFT-based method which yields a much more realistic prediction for the RI. Moreover, this method allows us to get both the in-plane and out-of-plane complex RI of \hfds\ for a very long wavelength range. We present the results in Figure \ref{fig:RI_BG}(c), where the dashed lines represent the predicted theoretical data. We observe that indeed the theoretical extinction coefficient of both in-plane (dashed green) and out-of-plane (dashed grey) components are negligible above 540 nm. As expected, we also see that the theoretical in-plane RI (dashed red) stays above 3 for the whole visible range, even reaching values of 4.5 in the blue spectral range. The theoretical out-of-plane RI (dashed blue) varies between 2 and 2.5 in the visible range. Overall, Figure \ref{fig:RI_BG} shows that the combination of DFT-coarse plus DFT-refined simulations is a high-throughput theoretical tool to discover new materials.

%\subsection{Refractive index measurements}
We now turn to an experimental demonstration of \hfds, where we first measure the complex refractive index tensor to verify the theoretical predictions. High-purity ($>99.995\%$) \hfds\ crystals are purchased from a commercial vendor and mechanically exfoliated to produce flakes that are transferred onto a silicon substrate. The crystallinity of the exfoliated flakes is verified with Raman spectroscopy measurements (see Supplementary Note~2). We measure the refractive index using an imaging ellipsometer (see Methods) and the experimental results (solid lines) are shown along with the theoretical results in Fig.~\ref{fig:RI_BG}(b). We observe that the agreement is very good in all the measurements except for the out-of-plane refractive index, where we observe discrepancies of the order of 10\%. Nonetheless, we experimentally verify our theoretical predictions, namely that \hfds\ is a high-refractive-index material with very low losses in the visible range.

\subsection{Chemical instability}
\begin{figure}
    \centering
    \includegraphics[width=0.95\textwidth]{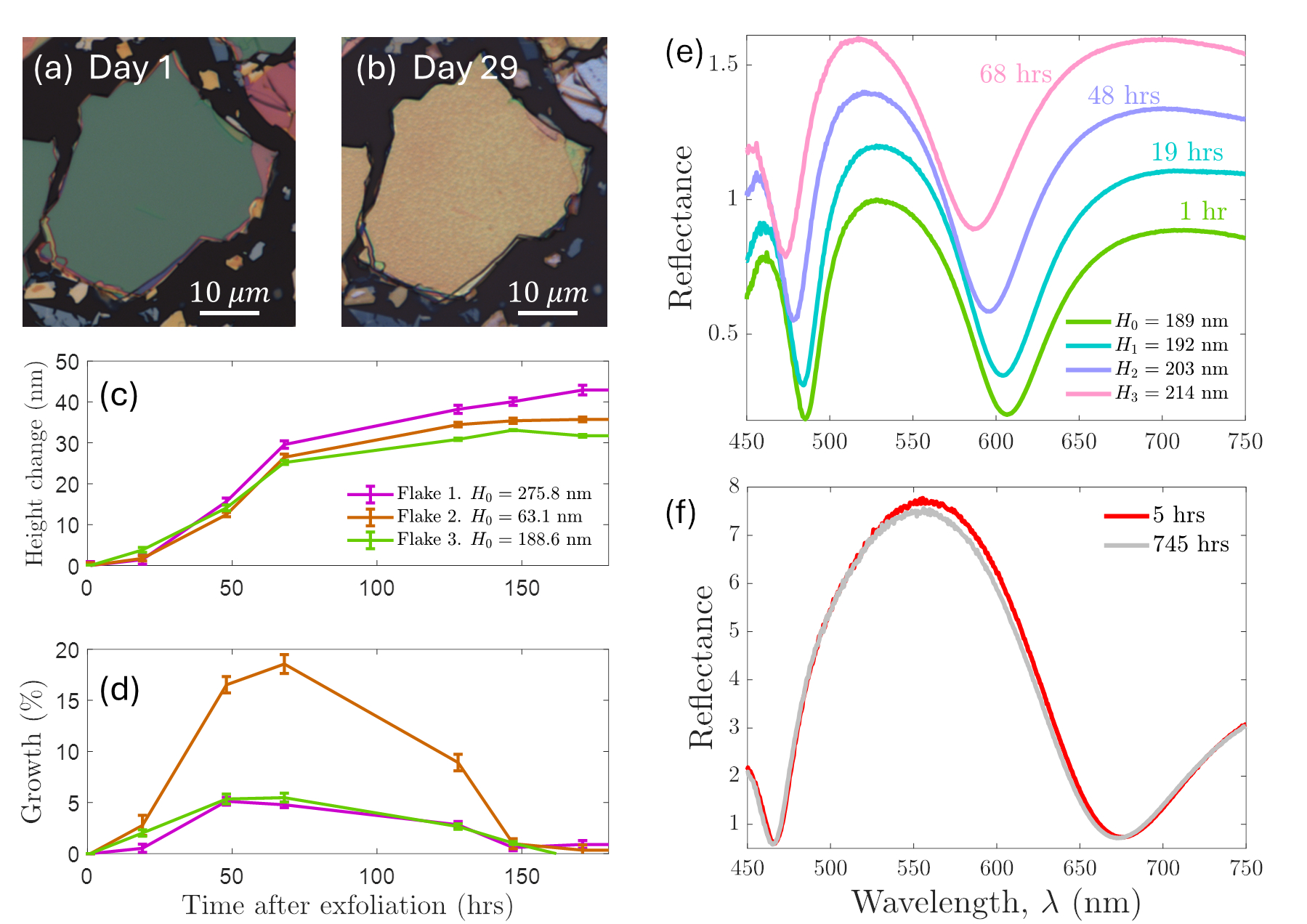}
    \caption{\textbf{Mitigating the chemical instability of HfS$_2$.} (a) Optical image of a \hfds\ flake right after exfoliation. (b) Optical image of the same \hfds\ flake after being exposed to ambient conditions for 29 days. Colour and structural (bubbles) changes are observed. (c) Height change (in nm) as a function of time (in hrs) for three different \hfds\ flakes. Their initial heights are 275.8, 63.1, and 188.6 nm respectively. (d) Growth (in \%) as a function of time after exfoliation (in hrs). The growth is computed as increment of height over the current height. (e) Reflectance as a function of the wavelength and time after exfoliation for the flake with an initial height $H_0=188.6$ nm. (f) Comparative reflectance of a \hfds\ flake 5 hrs and 745 hrs after it was exfoliated. The flake is kept in a desiccator with a humidity reduced to 10\%. The data is normalized with respect to the substrate.}
    \label{fig:oxidation_study}
\end{figure}

We have observed that the \hfds\ flakes are chemically unstable at ambient conditions under regular lab conditions (Fig.~\ref{fig:oxidation_study}(a-b)), consistent with previous findings~\cite{Chae2016,Mirabelli2016,Lai2018}. Prior studies have also demonstrated that \hfds\ can be intentionally converted into hafnia (HfO$_2$), resulting in a change in thickness~\cite{Lai2018, Jin2021}. Because we aim to use \hfds\ to fabricate nanoscale resonators for optical applications under standard room temperature and humidity conditions, we investigated the chemical stability of bulk \hfds\ flakes. For this study, we prepare exfoliated \hfds\ flakes and measure the initial thickness of three different flakes using atomic force microscopy (AFM). After the initial characterization, we leave the chip at ambient conditions and perform seven new thickness measurements for each flake over the course of 171 hours. 
The absolute accumulated growth and the relative daily growth of the flakes are shown in Fig.~\ref{fig:oxidation_study}(c-d) respectively. We observe that the three flakes exhibit similar behavior, with rapid growth during the first three days, followed by a plateau in daily growth approaching 0\% after six days. Due to the similar absolute growth, the thinnest flake has the highest daily relative growth, with values of almost 20\% in the third day. The thinnest flake also shows the highest accumulated growth of approximately 56\%, while the growth of the two thicker flakes stays in the range of 15-17\%. These values are well below the observations of Ref.~\cite{Chae2016}, where an accumulated growth of over 250\% was observed for a few-layered \hfds\ sample. 

%The exact growth mechanism of \hfds\ is still debated, where among other a self-limited oxidation process has been proposed, where the HfO$_2$ layer formed by oxygen reacting with the \hfds\ surface becomes thick enough to limit further diffusion of oxygen molecules~\cite{Jin2021}. However, it has also been suggested that water intercalation plays role, as reported in \cite{Lai2018}. 

In addition to thickness changes, we also tracked changes in reflectivity of one of the flakes during the first three days (Fig.~\ref{fig:oxidation_study}(e)). The reflectance spectra show pronounced dips due to the excitation of Fabry--Pérot resonances. We observe that the increase in flake thickness is associated with a blueshift of approximately 20~nm of the Fabry--Pérot resonances. This might seem a bit counter-intuitive, as the Fabry--Pérot resonance wavelength $\lambda_q$ follows the relation
\begin{equation}
    \lambda_q = \dfrac{2nH}{q},  
    \label{eq:FP}
\end{equation}
where $q$ is the mode order, $n$ is the refractive index, and $H$ is the distance between the reflective surfaces of the cavity. In our system, this distance corresponds to the flake thickness. By examining Eq.~(\ref{eq:FP}), we observe that Fabry--Pérot resonances should redshift when $H$ increases. However, another possibility exists: as the flakes undergo a chemical transformation, their refractive index may be affected. Notably, HfO$_2$ has a lower refractive index than \hfds~\cite{Polyanskiy2024}. We therefore propose that the observed blueshift is a result of the chemically-induced decrease in refractive index, which dominates over the increase in thickness. %We therefore propose that the observed blueshift is primarily due to a chemically induced decrease of the refractive index.

It is also worth noting that not only did we observe changes in the thickness and in the reflectivity of the \hfds\ flakes, but also a clear change in the structure of the material (Fig.~\ref{fig:oxidation_study}(a-b)). It is seen that the color of the flake changes, and that bubbles form on the surface. Similar bubble formation has been reported in previous works~\cite{Chae2016,Jin2021} and is believed to result from defect points around which oxidation occurs. The exact mechanism underlying the chemical instability of \hfds\ remains debated. Proposed explanations include a self-limited oxidation process, in which a HfO$_2$ layer forms through the reaction of oxygen with the \hfds\ surface, eventually limiting further oxygen diffusion~\cite{Jin2021}. Alternatively, water intercalation has also been suggested to play a role~\cite{Lai2018}. Investigating these mechanisms in detail is beyond the scope of this work. Instead, we adopt a pragmatic approach to mitigate the chemical instability of \hfds\ by exploring different environmental conditions.

To mitigate the chemical instability, we investigated long-term storage of \hfds\ flakes in both inert argon atmosphere in a glovebox as well as a simpler humidity-controlled desiccator. We find that both storage conditions effectively stabilize the material. In particular, we track the reflectance for a period of one month for two samples, where one is stored in ambient conditions, and the other one is stored in a desiccator which reduces the humidity down to 10\%. A subset of these results shows that the reflectance remains virtually unchanged (Fig.~3(f)), with detailed findings provided in Supplementary Note~3. These findings demonstrate a strategy for mitigating the chemically induced changes and retaining the desired optical properties of \hfds\ by storing the material in a controlled atmosphere.

\subsection{Fabrication of optical HfS$_2$ nanodisks}
\begin{figure}
    \centering
    \includegraphics[width=1\textwidth]{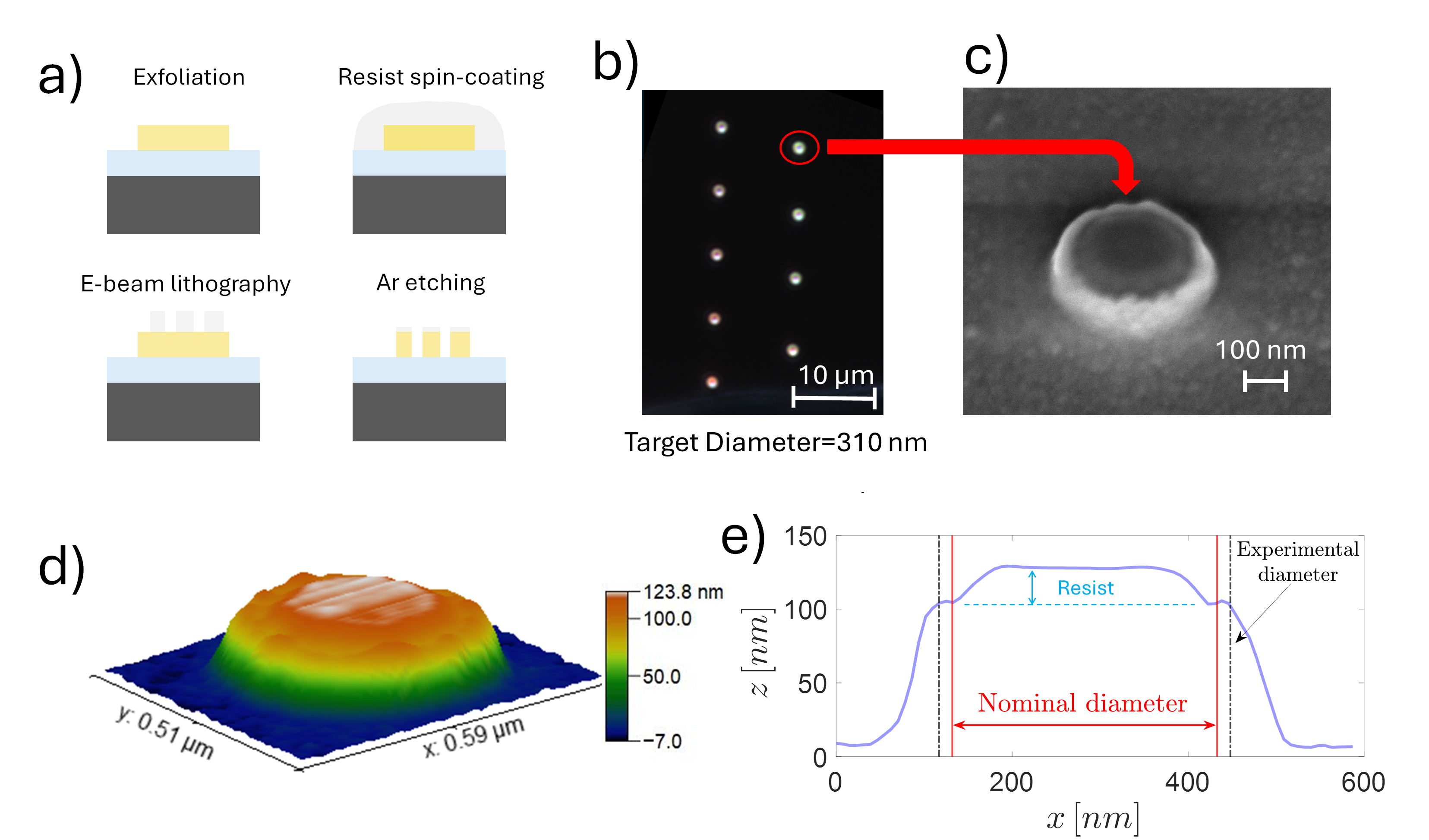}
    \caption{\textbf{Fabrication of optical HfS$_2$ resonators.} (a) The fabrication process consists of exfoliation, resist spin-coating, electron-beam lithography and argon etching. (b) Dark-field optical image at a 100x magnification of a series of \hfds\ nanoresonators with nominal diameters ranging from 260 to 350~nm in steps of 10~nm. A nanoresonator with a nominal diameter of 310~nm is encircled in red and imaged with scanning electron microscopy in (c). (d) Three-dimensional plot of a height measurement using an AFM. Both (c-d) reveal residual resist on top of the nanoresonator and redeposition of \hfds\ on the sidewalls. (e) AFM height measurement along a transverse line for the same resonator as (d), indicating the nominal diameter of 310~nm. The residual resist thickness and actual diameter are also indicated.}
    \label{fig:HfS2_fab}
\end{figure}

Our goal is to leverage the high refractive index and low losses of \hfds\ to create nanostructures that support optical resonances in the visible range. This work represents, to the best of our knowledge, the first demonstration of lithographic patterning of bulk \hfds\ flakes. We fabricate 100~nm thick nanodisks with diameters from 100 to 350~nm, following the procedure schematically shown in Fig.~\ref{fig:HfS2_fab}(a). To begin, \hfds\ is exfoliated onto a silicon chip with a 90~nm silica layer, and 100~nm thick flakes are selected via AFM. The flakes are spin-coated with a negative electron-beam resist and then patterned via e-beam lithography to define arrays of nanodisks with the target diameters (see Methods for details).

Developing a dry etching recipe suitable for \hfds\ was a key fabrication challenge, as no prior methods have been reported. We initially tested an SF$_6$-based etching process commonly used for other TMDCs~\cite{Danielsen2021,Sarbajna2024}, but it turned out to yield a negligible etch rate. We therefore resorted to a physical etching approach using argon sputtering and optimized the etching duration to fully pattern 100~nm thick flakes. For thicker flakes, the etching was partial, creating a different set of nanostructures (see Supplementary Note~4).

A dark-field optical image (at 100x magnification) of some of our fabricated \hfds\ nanodisks are shown in Fig.~\ref{fig:HfS2_fab}(b). In it, we observe nanodisks with nominal diameters ranging from 260 to 350~nm in steps of 10~nm. For convenience, the rest of the nanodisks, with nominal diameters ranging from 100 to 260~nm in steps of 10~nm, are not shown. An unintended consequence of our etching procedure is material re-deposition, where \hfds\ molecules removed from the flake redeposit on the sidewalls of the resist. As a result, the fabricated nanodisks exhibit slanted sidewalls rather than straight, vertical ones, with nominal diameters approximately retained at the top of each disk but widened at the base (Fig.~\ref{fig:HfS2_fab}(c-e)). The AFM and scanning electron microscopy images also reveal a residual thin resist layer on top of the nanodisks, and that the diameter of the fabricated nanodisks is approximately 30~nm larger than the nominal diameter (as measured at the top surface). We have checked the difference between the nominal and real diameters of all the nanodisks, and consistently found that the real diameters are approximately 30~nm wider.

\subsection{Resonant light scattering from HfS$_2$ nanodisks}
\begin{figure}
    \centering
    \includegraphics[width=0.72\textwidth]{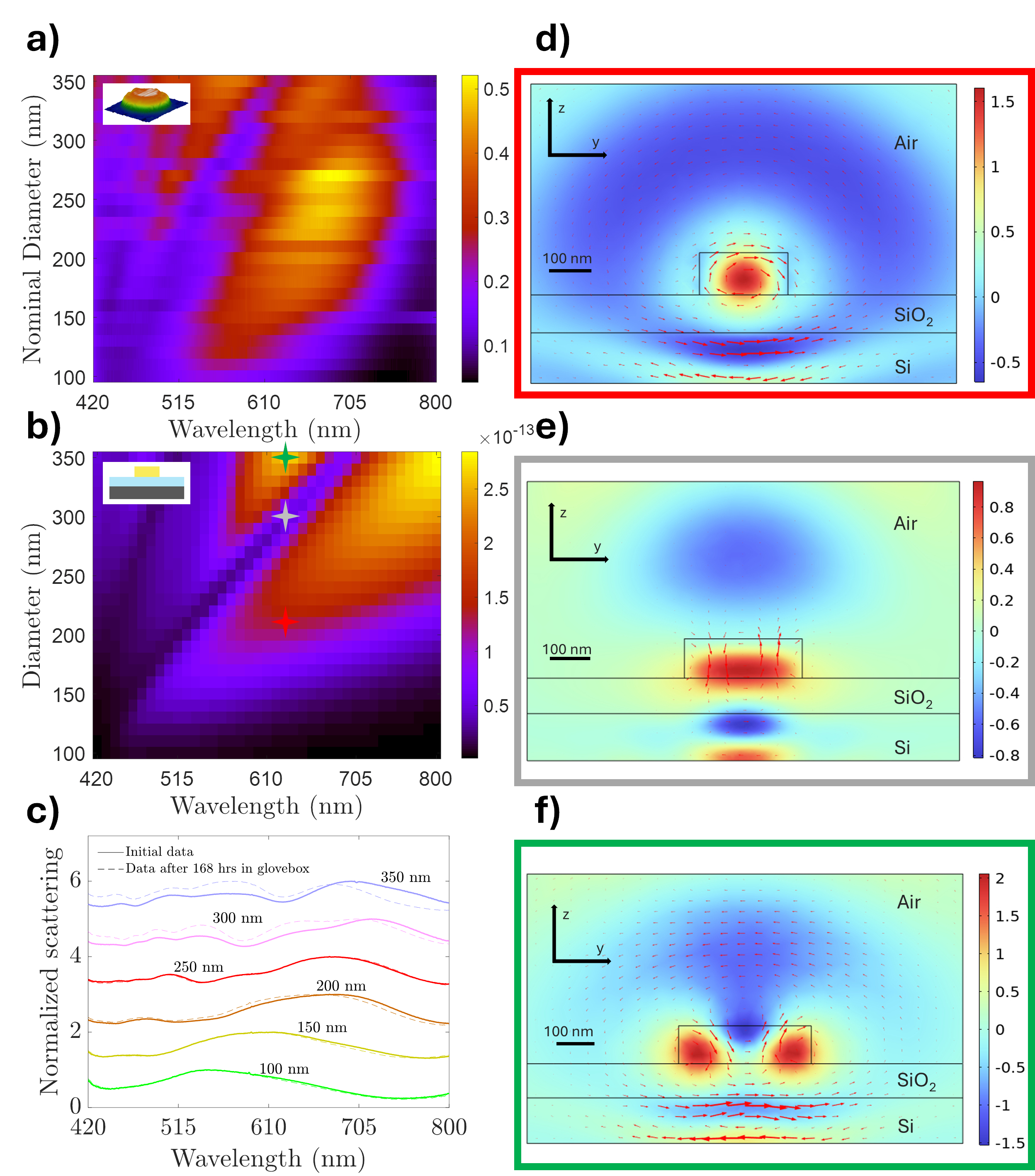}
    \caption{\textbf{Mie resonances in HfS$_2$ nanodisks.} (a) Dark-field scattering measurements of the fabricated nanodisks (see inset). Each row of the plot corresponds to the spectral scattering measurement of a single resonator with a different nominal diameter. All of the scattering measurements are normalized by the same background signal. (b) Numerical scattering cross section calculated for a \hfds\ disk with a varying diameter. The disk is placed on a 90~nm thick layer of silica on semi-infinite silicon (see inset). The colored-stars refer to three specific diameters, whose electromagnetic fields are plotted in (d-f) for $\lambda=630$~nm. (c) Normalized dark-field scattering measurements for six nanodisks with varying diameters. The zero scattering line of each plot is moved up one unit. The solid lines are measured right after the resonators are etched, while the dashed lines are measurements after the sample has been in a glovebox for 168 hrs, with a total of six hour subjected to ambient condition. (d-f) Electromagnetic scattered field at a wavelength of $\lambda=630$ nm for diameters $d=210$~nm, $d=300$~nm, and $d=350$~nm, respectively. The boxes are color-coded to match the colored stars in (b). The real part of the electric field in the $\xhat$ polarization is plotted as a colour map, and the magnetic field is plotted as red arrows.}
    \label{fig:results}
\end{figure}

%\subsection{Experimental results}
We conduct dark-field spectroscopy measurements on the \hfds\ nanodisks to assess their ability to support optical resonances (see Methods). We minimize the exposure of the nanostructures to ambient conditions by careful planning of all our experiments and storing our samples in an argon environment whenever possible. The scattering spectra of all the fabricated nanodisks sorted as a function of their nominal diameter $d$ are shown in Fig.~\ref{fig:results}(a). As previously mentioned, the real diameter is about 30~nm wider than the nominal diameter. All of the spectra are normalized by the same spectrum, collected from a white powder, allowing us to see how the intensity of the scattering peaks grows or gets reduced as we change the diameter of the nanodisks (see Methods). The spectra show a spectral local minima forming around $d = 170 $ nm, which red-shifts with increasing diameters. We also observe two very well defined peak families. One of the peak families is broader than the other one, and we observe that even the smallest resonators of $d=100$~nm can host this resonance at around a wavelength of $\lambda = 550$~nm. This resonance redshifts for larger diameters sizes, and it also increases in intensity. The other resonance family presents sharper peaks at shorter wavelengths. The trend begins at around $d = 200$ nm for $\lambda = 480$~nm and it slowly redshifts with increasing diameters, reaching a resonance wavelength of $\lambda=565$~nm for a diameter of $d=350$~nm. The redshifting of the scattering peaks for increasing resonator sizes is typically what one expects for Mie resonances. 

%\subsection{Numerical simulations}
To understand the nature of these spectral features, we conduct 3D full-field electromagnetic simulations using COMSOL Multiphysics, which solves Maxwell's equations using the finite element method. We simulate the HfS$_2$ resonators as 100~nm thick disks on top of a two layer substrate, consisting of a 90~nm thick layer of SiO$_2$ on top of a semi-infinite silicon substrate. The system is excited with a normally-incident plane wave, linearly-polarized in $\widehat{\mathbf{x}}$ (see Methods). For the refractive index of \hfds, we have used our theoretical data. The computed scattering cross section (Fig.~\ref{fig:results}(b)) shows very similar trends to the experimental measurements, despite (i) the different illumination conditions - normally incident plane wave in simulations vs. tightly-focused Bessel beam in dark-field experiment; and (ii) the idealized modeling of the HfS$_2$ nanoresonators as perfectly shaped nanodisks vs. nanodisks with slanted walls shown in Fig. \ref{fig:HfS2_fab}(d-e). In particular, we observe a pronounced scattering peak that forms around a wavelength of 500~nm for the smaller resonators, which redshifts as the diameter increases. The nature of this resonance can be identified by inspecting the scattered electromagnetic field, which reveals a field pattern of intense electric field in the center of the nanodisk with a magnetic field spiraling around this center (Fig.~\ref{fig:results}(d)). Based on the field profile, we interpret this resonance as due to the excitation of an electric dipole Mie resonance. In addition to this prominent dipolar peak, we also observe that nanodisks with diameters greater than 210~nm yield a second scattering peak at shorter wavelengths. The electromagnetic field associated to this scattering peak shows a pattern with two electric field lobes at the lateral surfaces of the resonator, each of them associated with a spiraling magnetic field in-phase with respect to the other (Fig.~\ref{fig:results}(f)). Furthermore, there is another maximum at the top surface. Given the shape of the electromagnetic field, we infer that the peak is due to a higher multipolar order, possibly a quadrupolar resonance. Notice that these two resonances are also present in the experimental results. Finally, another feature that is also present in both the experiments and simulations is the spectral minima which separates the dipolar and quadrupolar resonances. The difference between the theoretical plot and the numerical one is that the numerical plot yields a minima line which is almost linear, whereas our experimental results yield a minima line which is curved. Moreover, the experimental minima line is not as well defined as the numerical one. The field structure of this scattering minimum is depicted in Fig. \ref{fig:results}(e) and resembles that of a directional forward-scatterer or photonic nanojet \cite{AllenThesis2014,Darafsheh2021,Surdo2021}. Overall, we can see that our simulations capture all the prominent features of the experimental measurements. 

%\subsection{Stability of the optical response}
To conclude our optical study, we compared the dark-field optical spectra of the nanodisks right after fabrication, as well as seven days later. During these seven days, the sample was stored in an argon-rich atmosphere in a glovebox. Apart from the final spectroscopy itself, the samples were only taken out for AFM measurements. The nanodisks have been exposed to ambient conditions for approximately six hours. In Fig.~\ref{fig:results}(c), we display dark-field spectra of selected nanodisks obtained right after the fabrication (solid lines), and we compare them with the spectra of the same nanodisks seven days later (dashed lines). The plot contains spectra from structures within a wide range of nominal diameters in order to give a full picture of which sizes are affected the most by exposure to ambient conditions. Each spectrum is normalized to its maximum intensity value for better comparison. This comparison shows that the nanodisks with diameters below 200 nm yield almost the same optical response after seven days, while the larger nanodisks show moderate changes due to the exposure to ambient conditions. It is worth noting that the spectral response of nanoresonators is much more sensitive to size or refractive index changes than that of bulk flakes. Thus, overall we conclude that keeping the nanoresonators in an argon-rich glovebox is a simple and effective way to stabilize the spectral response of the \hfds\ nanodisks. 

\section{Discussion}
This work demonstrates a rational approach to identifying and evaluating high-refractive-index van der Waals materials by combining high-throughput DFT screening with experimental validation. The screening methodology not only highlights \hfds\ as a promising candidate for visible-spectrum photonics but also provides a comparative overview of numerous materials with potential applications across different spectral ranges. Notably, the computational framework identifies many materials with super-Mossian behavior at lower photon energies, which could warrant further experimental exploration to expand the landscape of viable high-index dielectrics.

Focusing on \hfds, our study establishes it as an intriguing material for visible photonics, offering a high in-plane refractive index ($>3$), low optical losses, and pronounced anisotropy. These properties are leveraged to experimentally demonstrate its suitability for supporting Mie resonances in nanodisk resonators. However, the chemical instability of \hfds\ under ambient conditions remains a challenge, that we mitigated by storing the material in argon-rich or low-humidity environments. Another approach could be to encapsulate \hfds\ in protective coatings, such as resist or other two-dimensional materials like hexagonal boron nitride. These approaches could enhance the long-term stability of \hfds\ and expand its practical applicability. Future work should also address optimizing the fabrication process for \hfds\--based photonic devices. While the physical etching approach used in this study successfully produced functional nanodisks, it introduced redeposition effects that could be mitigated by developing a chemical etching technique. Such improvements would enable more precise patterning and reduce fabrication-related imperfections, further unlocking the material’s potential.

Beyond \hfds, this study underscores the versatility of the high-throughput computational screening approach in identifying high-index optical materials. Expanding the screening to include ternary and quaternary compounds could uncover even more candidates with tailored optical and electronic properties. This methodology, which combines predictive modeling with targeted experimental validation, provides a robust framework for the rational discovery of improved optical materials. By streamlining the identification process and ensuring high-fidelity predictions, this approach can accelerate the development of new photonic materials for a wide range of technological applications.

\section{Methods}
\subsection{High throughput screening}
%All DFT simulations are performed using the GPAW electronic structure code~\cite{Mortensen2024,yan2011linear}.
The refractive index dataset was generated using the method detailed in our previous paper in Ref. \cite{Svendsen2022}. For completeness, we reproduce the details here:

\textbf{Electronic Structure:} All ground- and excited state calculations were performed with ASE and the GPAW electronic structure code \cite{Larsen2017,Mortensen2024}. The atomic structure and the unit cell of the materials were relaxed until the maximum force (stress) was below $10^{-4}$~eV Å$^{-1}$ (0.002~eV Å$^{-3}$). The PBE functional for exchange and correlation effects, a $\Gamma-$point centered $k$-point grid with a density of 6.0~1/Å$^{-1}$ along each direction in the Brillouin zone, an 800~eV plane wave cutoff, and a Fermi-Dirac smearing of 50~meV was used. Van der Waals interactions were taken into account by the D3 correction scheme. For the most promising materials, we performed G$_0$W$_0$ calculations using the asr.gw recipe to determine their proper electronic bandgap~\cite{Gjerding2021}.

\textbf{RPA Calculations for the screening:} The optical permittivity, $\epsilon(\omega)$, was calculated within the RPA using the dielectric function module in GPAW. From $\epsilon(\omega)$, the refractive index and extinction coefficient were calculated as $\mathrm{Re} \left[  \sqrt{\epsilon(\omega)} \right]$ and  $\mathrm{Im} \left[  \sqrt{\epsilon(\omega)} \right]$, respectively. To ensure convergence across all materials, a $k$-point grid with a high density of 20.0~1/Å$^{-1}$ along each direction in the Brillouin zone was employed and conduction bands up to five times the number of valence bands were included. The calculations were performed on a nonlinear frequency grid with an initial frequency spacing of 0.5~meV, a broadening of 50~meV, and a local field cutoff of 50~eV.

\textbf{BSE+ calculations:} The BSE+ method seamlessly combines a BSE description of the low-energy transitions with an RPA treatment of higher-energy transitions. As described in detail in Ref. \cite{Søndersted2024}, this is essential to obtain a quantitatively accurate description of refractive index in systems where excitonic effects are important. Because the PBE exchange correlation functional is known to underestimate electronic band gaps, for the BSE+ calculations of \hfds\ we apply a scissors operator to adjust the band gaps, aligning the lowest peak observed in the refractive index calculated with BSE with the lowest peak in the refractive index observed in the experimental data. The RPA calculation was performed using 130 bands, with a local field cutoff of 80~eV. This was based on a ground state calculation performed using the PBE functional, a plane wave cutoff of 800~eV and a $\Gamma$-point centered Monkhorst-Pack 36x36x22 $k$-point grid. The BSE calculation was performed using 4 valence bands and 3 conduction bands, a local field cutoff of 80~eV and a broadening of 0.1~eV. The electronic screening was calculated with 60 electronic bands. For the calculation of the polarizabilities in the in- and out-of-plane direction $\Gamma$-point centered Monkhorst-Pack grids of 12x12x6 and 12x12x8 were used respectively.

\subsection{Imaging ellipsometry}
The refractive index and extinction coefficient measurements have been done with an imaging ellipsometer Accurion EP4. We performed measurements over a broad spectral range from 360~nm to 1700~nm in 1~nm steps for two incident angles: $45^{\circ}$ and $50^{\circ}$. For the optical modeling of ellipsometry, we used the Tauc-Lorentz oscillator approach for both in-plane and out-of-plane optical constants \cite{Ermolaev2021}.

\subsection{Nanofabrication details}
%First, using mechanical exfoliation, we have transferred some \hfds\ material from a crystal onto a chip made of silicon with a layer of 90 nm of silica on top. Our target is getting 100 nm thick flakes, so we perform a number of exfoliations such that we can statistically easily obtain such thicknesses. Still, mechanical exfoliation does not allow for a precise control of thicknesses, so we use an AFM to find the flakes whose thickness is $(100 \pm 10)$ nm. The second step is to spin-coat the entire sample with a negative electron-beam resist ARN-7520new, as seen in Fig. \ref{fig:HfS2_fab}(a). The resist will serve as a protective layer during the subsequent etching process. Then, using e-beam lithography, we expose the areas of the resist that we are interested in, creating arrays of nano-disks with varying diameters. Later on, we remove the unexposed resist (development) and obtain a nanostructure formed by resist-\hfds-SiO$_2$-Si as shown in Fig. \ref{fig:HfS2_fab}(a). The last fabrication step is to etch the material. We have chosen a physical etching method based on Ar-sputtering. The length of the etching has been chosen so that the 100 nm thick flakes were completely etched down. 
Thin flakes were mechanically exfoliated from a \hfds\ crystal from HQ Graphene onto a 90~nm SiO$_2$ layer on top of a Si substrate. Our target was getting 100~nm thick flakes, so we performed a number of exfoliations such that we could statistically easily obtain such thicknesses. Still, mechanical exfoliation does not allow for a precise control of thicknesses, so we used an AFM to find the flakes whose thickness is  $(100 \pm 10)$ nm.
For lithography, ARN-7520.new from Allresist was used as a negative electron-beam resist. The resist was spin-coated at 2000~rpm for 60~s and post-baked at 85$^{\circ}$~C for 60~s. Then, using electron-beam lithography, we defined the mask for the nano-disks in a 30kV Raith eLINE Plus system using a dose of 50~$\mu$Ccm$^{-2}$ and an aperture of 30~$\mu$m. Next, the resist was developed in an AR300-47 solution from Allresist for 70~s. Finally, the pillars were etched by Ar-sputtering in an IBE/IBDS Ionfab 300 set up with a 20~mA acceleration current at a 30$^{\circ}$ incidence angle for 4 minutes with a flow rate of 10~sccm. The length of the etching was chosen so that the 100~nm thick flakes were completely etched down, which required preliminary experiments to estimate the etch rate of Ar in each material. By fixating the samples in the etch chamber with Kapton tape, which is unaffected by the physical etch, its location could serve as a reference when measuring the post-etch height profile on the DekTak 150 profiler. The etch rate of Ar sputtering in \hfds\ found with this method was $28.0 \pm 2.2$~nm/min.

\subsection{Atomic Force Microscopy}
The thickness of the flakes and height profiles of the fabricated nanostructures were measured by a Dimension Icon-PT AFM from Bruker AXS in tapping mode. A scan rate of 0.5~Hz was used with at least 128 data points in each line scanned, ensuring that abrupt height changes were well resolved.

\subsection{Microscopy and optical spectroscopy}
The bright and dark field microscopy were performed using a Nikon Eclipse LV100ND microscope and an OSL-2 fiber-coupled unpolarized halogen light source for top illumination. An Andor Kymera 328i spectrograph with a slit width of 200~$\mu$m and a central wavelength of 650~nm was used to perform the reflectance and scattering spectra from the flat flakes and the nano-disks, respectively. Reflectance spectra were acquired with a 20x objective (Nikon, NA$=0.45$). The measured spectra were normalized by the reflection spectra of a protected silver mirror (Thorlabs, PF10-03-P01), which was exposed to the same illumination. For dark-field measurements, we used a 100x objective (Nikon, NA$=0.9$) to ensure maximum collection of scattered light. The disks were strategically fabricated with sufficient spacing to eliminate noise contributions from adjacent structures, and were exposed long enough to have a high signal-to-noise ratio. The measured spectra were normalized by the signal collected from a Lambertian scatterer exposed to the same conditions as the sample. The dark counts (pixel values when no optical signal is present) of the detector were also recorded and subtracted prior to normalization.

\subsection{Electromagnetic simulations}
We simulate the scattering response of the \hfds\ nano-disks using Comsol Multiphysics. A cross-section of the geometry of our system can be observed in Fig. \ref{fig:results}(d-f). That is, we place the 100~nm thick nano-disks on top of a 90~nm silica layer, which lays on top of a semi-infinite silicon layer. The nano-disks are surrounded by a semi-infinite layer of air. To properly model the semi-infinite air and silicon layers, we place PML layers on top of the air, and at the bottom of the silicon layer. Moreover, we choose a material thickness of $\lambda/2n$, with $n$ being the refractive index of the material in consideration. To be on the safe side, we implement a thickness of 400~nm for the air, and a thickness of 120~nm for the silicon layer. Notice that the 400~nm of air are applied from the top of the nano-disk, which gives an air thickness of 500~nm from the silica top surface. Similar considerations are done in the transverse direction. That is, we also place PML layers on the sides of our simulations, and we make the width of the simulation to be 800~nm (one full wavelength) plus the diameter of the nano-disk - i.e., we leave $\lambda/2$ from both sides of the nano-disk walls. Even though silica and silicon have greater refractive indices, we keep the same thickness as the air. All the PMLs that we place have a thickness of $\lambda_{\mathrm{max}}/3$. We run simulations for a range of diameters varying from 100 to 350~nm in 10~nm steps. The wavelength has also been varied from 420 to 800~nm in 10~nm steps. We have computed the scattered cross section $C^\textrm{sca}$  on the particle by computing the flow of the Poynting vector across the surface of the particle which is in contact with the air domain: 
\begin{equation}
C^\textrm{sca}=\dfrac{\displaystyle \int_S  \mathbf{P}^\textrm{sca} \cdot \mathbf{\mathrm{d}S}} {I_0} 
\end{equation}
with $\mathbf{P}^\textrm{sca}$ being the Poynting vector associated to the scattered field, and $I_0$ being the intensity of the incoming plane wave.

\section{Acknowledgments}
X.Z.-P. thanks Ganesh Ghimire for helping in exfoliating the samples. X.Z.-P. and S.R. acknowledge funding from the Villum Foundation (VIL50376). K.S.T. is a Villum Investigator supported by VILLUM FONDEN (grant no. 37789). We acknowledge support from the Novo Nordisk Foundation Data Science Research Infrastructure 2022 Grant:  A high-performance computing infrastructure for data-driven research on sustainable energy materials, Grant no. NNF22OC0078009. 

\section{Author contributions}
M.K.S. and A.H.S have carried out the high throughput screening of materials. A.S., J.P.S. and A.L.R. have fabricated the \hfds\ nanodisks. J.P.S. and A.L.R. have taken the AFM images of the nanodisks. X.Z.P. has taken the SEM images. J.P.S. and A.L.R. have recorded the optical spectra. J.P.S., A.L.R., and X.Z.-P. have done the two separate oxidation studies. G.E. has measured the refractive index of \hfds. G.T. has done the Raman study. T.M.B. created the optical materials database. X.Z.-P. has carried out the numerical simulations. X.Z.-P. has made the figures. X.Z.-P. and S.R. have written the manuscript with the help of all the rest of authors. V.S.V., K.S.T., and S.R. have supervised the worked. 

% Bibliography
%\bibliography{biblio}
%\bibliography{../../Bibliography/biblio}

\begin{thebibliography}{45}%
\makeatletter
\providecommand \@ifxundefined [1]{%
 \@ifx{#1\undefined}
}%
\providecommand \@ifnum [1]{%
 \ifnum #1\expandafter \@firstoftwo
 \else \expandafter \@secondoftwo
 \fi
}%
\providecommand \@ifx [1]{%
 \ifx #1\expandafter \@firstoftwo
 \else \expandafter \@secondoftwo
 \fi
}%
\providecommand \natexlab [1]{#1}%
\providecommand \enquote  [1]{``#1''}%
\providecommand \bibnamefont  [1]{#1}%
\providecommand \bibfnamefont [1]{#1}%
\providecommand \citenamefont [1]{#1}%
\providecommand \href@noop [0]{\@secondoftwo}%
\providecommand \href [0]{\begingroup \@sanitize@url \@href}%
\providecommand \@href[1]{\@@startlink{#1}\@@href}%
\providecommand \@@href[1]{\endgroup#1\@@endlink}%
\providecommand \@sanitize@url [0]{\catcode `\\12\catcode `\$12\catcode
  `\&12\catcode `\#12\catcode `\^12\catcode `\_12\catcode `\%12\relax}%
\providecommand \@@startlink[1]{}%
\providecommand \@@endlink[0]{}%
\providecommand \url  [0]{\begingroup\@sanitize@url \@url }%
\providecommand \@url [1]{\endgroup\@href {#1}{\urlprefix }}%
\providecommand \urlprefix  [0]{URL }%
\providecommand \Eprint [0]{\href }%
\providecommand \doibase [0]{https://doi.org/}%
\providecommand \selectlanguage [0]{\@gobble}%
\providecommand \bibinfo  [0]{\@secondoftwo}%
\providecommand \bibfield  [0]{\@secondoftwo}%
\providecommand \translation [1]{[#1]}%
\providecommand \BibitemOpen [0]{}%
\providecommand \bibitemStop [0]{}%
\providecommand \bibitemNoStop [0]{.\EOS\space}%
\providecommand \EOS [0]{\spacefactor3000\relax}%
\providecommand \BibitemShut  [1]{\csname bibitem#1\endcsname}%
\let\auto@bib@innerbib\@empty
%</preamble>
\bibitem [{\citenamefont {Bayati}\ \emph {et~al.}(2019)\citenamefont {Bayati},
  \citenamefont {Zhan}, \citenamefont {Colburn}, \citenamefont
  {Zhelyeznyakov},\ and\ \citenamefont {Majumdar}}]{Bayati2019}%
  \BibitemOpen
  \bibfield  {author} {\bibinfo {author} {\bibfnamefont {E.}~\bibnamefont
  {Bayati}}, \bibinfo {author} {\bibfnamefont {A.}~\bibnamefont {Zhan}},
  \bibinfo {author} {\bibfnamefont {S.}~\bibnamefont {Colburn}}, \bibinfo
  {author} {\bibfnamefont {M.~V.}\ \bibnamefont {Zhelyeznyakov}},\ and\
  \bibinfo {author} {\bibfnamefont {A.}~\bibnamefont {Majumdar}},\ }\bibfield
  {title} {\bibinfo {title} {{Role of refractive index in metalens
  performance}},\ }\href {https://doi.org/10.1364/AO.58.001460} {\bibfield
  {journal} {\bibinfo  {journal} {Appl. Opt.}\ }\textbf {\bibinfo {volume}
  {58}},\ \bibinfo {pages} {1460} (\bibinfo {year} {2019})}\BibitemShut
  {NoStop}%
\bibitem [{\citenamefont {Yang}\ and\ \citenamefont {Fan}(2017)}]{Yang2017}%
  \BibitemOpen
  \bibfield  {author} {\bibinfo {author} {\bibfnamefont {J.}~\bibnamefont
  {Yang}}\ and\ \bibinfo {author} {\bibfnamefont {J.~A.}\ \bibnamefont {Fan}},\
  }\bibfield  {title} {\bibinfo {title} {{Analysis of material selection on
  dielectric metasurface performance}},\ }\href
  {https://doi.org/10.1364/OE.25.023899} {\bibfield  {journal} {\bibinfo
  {journal} {Opt. Express}\ }\textbf {\bibinfo {volume} {25}},\ \bibinfo
  {pages} {23899} (\bibinfo {year} {2017})}\BibitemShut {NoStop}%
\bibitem [{\citenamefont {Rybin}\ \emph {et~al.}(2017)\citenamefont {Rybin},
  \citenamefont {Koshelev}, \citenamefont {Sadrieva}, \citenamefont {Samusev},
  \citenamefont {Bogdanov}, \citenamefont {Limonov},\ and\ \citenamefont
  {Kivshar}}]{Rybin2017}%
  \BibitemOpen
  \bibfield  {author} {\bibinfo {author} {\bibfnamefont {M.~V.}\ \bibnamefont
  {Rybin}}, \bibinfo {author} {\bibfnamefont {K.~L.}\ \bibnamefont {Koshelev}},
  \bibinfo {author} {\bibfnamefont {Z.~F.}\ \bibnamefont {Sadrieva}}, \bibinfo
  {author} {\bibfnamefont {K.~B.}\ \bibnamefont {Samusev}}, \bibinfo {author}
  {\bibfnamefont {A.~A.}\ \bibnamefont {Bogdanov}}, \bibinfo {author}
  {\bibfnamefont {M.~F.}\ \bibnamefont {Limonov}},\ and\ \bibinfo {author}
  {\bibfnamefont {Y.~S.}\ \bibnamefont {Kivshar}},\ }\bibfield  {title}
  {\bibinfo {title} {High-q supercavity modes in subwavelength dielectric
  resonators},\ }\href@noop {} {\bibfield  {journal} {\bibinfo  {journal}
  {Physical review letters}\ }\textbf {\bibinfo {volume} {119}},\ \bibinfo
  {pages} {243901} (\bibinfo {year} {2017})}\BibitemShut {NoStop}%
\bibitem [{\citenamefont {Zambrana-Puyalto}\ and\ \citenamefont
  {Raza}(2024)}]{Zambrana2024}%
  \BibitemOpen
  \bibfield  {author} {\bibinfo {author} {\bibfnamefont {X.}~\bibnamefont
  {Zambrana-Puyalto}}\ and\ \bibinfo {author} {\bibfnamefont {S.}~\bibnamefont
  {Raza}},\ }\bibfield  {title} {\bibinfo {title} {Quality factor of dielectric
  spherical resonators},\ }\href {https://doi.org/10.1021/acsphotonics.4c00753}
  {\bibfield  {journal} {\bibinfo  {journal} {ACS Photonics}\ }\textbf
  {\bibinfo {volume} {11}},\ \bibinfo {pages} {3317} (\bibinfo {year}
  {2024})}\BibitemShut {NoStop}%
\bibitem [{\citenamefont {Boyd}(2008)}]{BoydBook}%
  \BibitemOpen
  \bibfield  {author} {\bibinfo {author} {\bibfnamefont {R.~W.}\ \bibnamefont
  {Boyd}},\ }\href@noop {} {\emph {\bibinfo {title} {Nonlinear Optics, Third
  Edition}}},\ \bibinfo {edition} {3rd}\ ed.\ (\bibinfo  {publisher} {Academic
  Press, Inc.},\ \bibinfo {address} {USA},\ \bibinfo {year} {2008})\BibitemShut
  {NoStop}%
\bibitem [{\citenamefont {Baranov}\ \emph {et~al.}(2017)\citenamefont
  {Baranov}, \citenamefont {Zuev}, \citenamefont {Lepeshov}, \citenamefont
  {Kotov}, \citenamefont {Krasnok}, \citenamefont {Evlyukhin},\ and\
  \citenamefont {Chichkov}}]{Baranov2017}%
  \BibitemOpen
  \bibfield  {author} {\bibinfo {author} {\bibfnamefont {D.~G.}\ \bibnamefont
  {Baranov}}, \bibinfo {author} {\bibfnamefont {D.~A.}\ \bibnamefont {Zuev}},
  \bibinfo {author} {\bibfnamefont {S.~I.}\ \bibnamefont {Lepeshov}}, \bibinfo
  {author} {\bibfnamefont {O.~V.}\ \bibnamefont {Kotov}}, \bibinfo {author}
  {\bibfnamefont {A.~E.}\ \bibnamefont {Krasnok}}, \bibinfo {author}
  {\bibfnamefont {A.~B.}\ \bibnamefont {Evlyukhin}},\ and\ \bibinfo {author}
  {\bibfnamefont {B.~N.}\ \bibnamefont {Chichkov}},\ }\bibfield  {title}
  {\bibinfo {title} {All-dielectric nanophotonics: the quest for better
  materials and fabrication techniques},\ }\href@noop {} {\bibfield  {journal}
  {\bibinfo  {journal} {Optica}\ }\textbf {\bibinfo {volume} {4}},\ \bibinfo
  {pages} {814} (\bibinfo {year} {2017})}\BibitemShut {NoStop}%
\bibitem [{\citenamefont {Yang}\ \emph {et~al.}(2023)\citenamefont {Yang},
  \citenamefont {Kang}, \citenamefont {Jung}, \citenamefont {Seong},
  \citenamefont {Jeon}, \citenamefont {Kim}, \citenamefont {Oh}, \citenamefont
  {Park}, \citenamefont {Kim},\ and\ \citenamefont {Rho}}]{Yang2023}%
  \BibitemOpen
  \bibfield  {author} {\bibinfo {author} {\bibfnamefont {Y.}~\bibnamefont
  {Yang}}, \bibinfo {author} {\bibfnamefont {H.}~\bibnamefont {Kang}}, \bibinfo
  {author} {\bibfnamefont {C.}~\bibnamefont {Jung}}, \bibinfo {author}
  {\bibfnamefont {J.}~\bibnamefont {Seong}}, \bibinfo {author} {\bibfnamefont
  {N.}~\bibnamefont {Jeon}}, \bibinfo {author} {\bibfnamefont {J.}~\bibnamefont
  {Kim}}, \bibinfo {author} {\bibfnamefont {D.~K.}\ \bibnamefont {Oh}},
  \bibinfo {author} {\bibfnamefont {J.}~\bibnamefont {Park}}, \bibinfo {author}
  {\bibfnamefont {H.}~\bibnamefont {Kim}},\ and\ \bibinfo {author}
  {\bibfnamefont {J.}~\bibnamefont {Rho}},\ }\bibfield  {title} {\bibinfo
  {title} {{Revisiting Optical Material Platforms for Efficient Linear and
  Nonlinear Dielectric Metasurfaces in the Ultraviolet, Visible, and
  Infrared}},\ }\href {https://doi.org/10.1021/acsphotonics.2c01341} {\bibfield
   {journal} {\bibinfo  {journal} {ACS Photonics}\ }\textbf {\bibinfo {volume}
  {10}},\ \bibinfo {pages} {307} (\bibinfo {year} {2023})}\BibitemShut
  {NoStop}%
\bibitem [{\citenamefont {Moss}(1950)}]{Moss1950}%
  \BibitemOpen
  \bibfield  {author} {\bibinfo {author} {\bibfnamefont {T.~S.}\ \bibnamefont
  {Moss}},\ }\bibfield  {title} {\bibinfo {title} {A relationship between the
  refractive index and the infra-red threshold of sensitivity for
  photoconductors},\ }\href@noop {} {\bibfield  {journal} {\bibinfo  {journal}
  {Proc. Phys. Soc. Sec. B}\ }\textbf {\bibinfo {volume} {63}},\ \bibinfo
  {pages} {167} (\bibinfo {year} {1950})}\BibitemShut {NoStop}%
\bibitem [{\citenamefont {Moss}(1985)}]{Moss1985}%
  \BibitemOpen
  \bibfield  {author} {\bibinfo {author} {\bibfnamefont {T.~S.}\ \bibnamefont
  {Moss}},\ }\bibfield  {title} {\bibinfo {title} {Relations between the
  refractive index and energy gap of semiconductors},\ }\href
  {https://doi.org/https://doi.org/10.1002/pssb.2221310202} {\bibfield
  {journal} {\bibinfo  {journal} {Phys. Status Solidi (B)}\ }\textbf {\bibinfo
  {volume} {131}},\ \bibinfo {pages} {415} (\bibinfo {year}
  {1985})}\BibitemShut {NoStop}%
\bibitem [{\citenamefont {Khurgin}(2022)}]{Khurgin2022}%
  \BibitemOpen
  \bibfield  {author} {\bibinfo {author} {\bibfnamefont {J.~B.}\ \bibnamefont
  {Khurgin}},\ }\bibfield  {title} {\bibinfo {title} {Expanding the photonic
  palette: Exploring high index materials},\ }\href@noop {} {\bibfield
  {journal} {\bibinfo  {journal} {Acs Photonics}\ }\textbf {\bibinfo {volume}
  {9}},\ \bibinfo {pages} {743} (\bibinfo {year} {2022})}\BibitemShut {NoStop}%
\bibitem [{\citenamefont {Ermolaev}\ \emph {et~al.}(2021)\citenamefont
  {Ermolaev}, \citenamefont {Grudinin}, \citenamefont {Stebunov}, \citenamefont
  {Voronin}, \citenamefont {Kravets}, \citenamefont {Duan}, \citenamefont
  {Mazitov}, \citenamefont {Tselikov}, \citenamefont {Bylinkin}, \citenamefont
  {Yakubovsky} \emph {et~al.}}]{Ermolaev2021}%
  \BibitemOpen
  \bibfield  {author} {\bibinfo {author} {\bibfnamefont {G.}~\bibnamefont
  {Ermolaev}}, \bibinfo {author} {\bibfnamefont {D.}~\bibnamefont {Grudinin}},
  \bibinfo {author} {\bibfnamefont {Y.}~\bibnamefont {Stebunov}}, \bibinfo
  {author} {\bibfnamefont {K.~V.}\ \bibnamefont {Voronin}}, \bibinfo {author}
  {\bibfnamefont {V.}~\bibnamefont {Kravets}}, \bibinfo {author} {\bibfnamefont
  {J.}~\bibnamefont {Duan}}, \bibinfo {author} {\bibfnamefont {A.}~\bibnamefont
  {Mazitov}}, \bibinfo {author} {\bibfnamefont {G.}~\bibnamefont {Tselikov}},
  \bibinfo {author} {\bibfnamefont {A.}~\bibnamefont {Bylinkin}}, \bibinfo
  {author} {\bibfnamefont {D.}~\bibnamefont {Yakubovsky}}, \emph {et~al.},\
  }\bibfield  {title} {\bibinfo {title} {Giant optical anisotropy in transition
  metal dichalcogenides for next-generation photonics},\ }\href@noop {}
  {\bibfield  {journal} {\bibinfo  {journal} {Nat. Commun.}\ }\textbf
  {\bibinfo {volume} {12}},\ \bibinfo {pages} {854} (\bibinfo {year}
  {2021})}\BibitemShut {NoStop}%
\bibitem [{\citenamefont {Verre}\ \emph {et~al.}(2019)\citenamefont {Verre},
  \citenamefont {Baranov}, \citenamefont {Munkhbat}, \citenamefont {Cuadra},
  \citenamefont {K{\"{a}}ll},\ and\ \citenamefont {Shegai}}]{Verre2019}%
  \BibitemOpen
  \bibfield  {author} {\bibinfo {author} {\bibfnamefont {R.}~\bibnamefont
  {Verre}}, \bibinfo {author} {\bibfnamefont {D.~G.}\ \bibnamefont {Baranov}},
  \bibinfo {author} {\bibfnamefont {B.}~\bibnamefont {Munkhbat}}, \bibinfo
  {author} {\bibfnamefont {J.}~\bibnamefont {Cuadra}}, \bibinfo {author}
  {\bibfnamefont {M.}~\bibnamefont {K{\"{a}}ll}},\ and\ \bibinfo {author}
  {\bibfnamefont {T.}~\bibnamefont {Shegai}},\ }\bibfield  {title} {\bibinfo
  {title} {{Transition metal dichalcogenide nanodisks as high-index dielectric
  Mie nanoresonators}},\ }\href {https://doi.org/10.1038/s41565-019-0442-x}
  {\bibfield  {journal} {\bibinfo  {journal} {Nat. Nanotechnol.}\ }\textbf
  {\bibinfo {volume} {14}},\ \bibinfo {pages} {679} (\bibinfo {year}
  {2019})}\BibitemShut {NoStop}%
\bibitem [{\citenamefont {Munkhbat}\ \emph {et~al.}(2022)\citenamefont
  {Munkhbat}, \citenamefont {Wr{\'{o}}bel}, \citenamefont {Antosiewicz},\ and\
  \citenamefont {Shegai}}]{Munkhbat2022}%
  \BibitemOpen
  \bibfield  {author} {\bibinfo {author} {\bibfnamefont {B.}~\bibnamefont
  {Munkhbat}}, \bibinfo {author} {\bibfnamefont {P.}~\bibnamefont
  {Wr{\'{o}}bel}}, \bibinfo {author} {\bibfnamefont {T.~J.}\ \bibnamefont
  {Antosiewicz}},\ and\ \bibinfo {author} {\bibfnamefont {T.~O.}\ \bibnamefont
  {Shegai}},\ }\bibfield  {title} {\bibinfo {title} {{Optical Constants of
  Several Multilayer Transition Metal Dichalcogenides Measured by Spectroscopic
  Ellipsometry in the 300–1700 nm Range: High Index, Anisotropy, and
  Hyperbolicity}},\ }\href {https://doi.org/10.1021/acsphotonics.2c00433}
  {\bibfield  {journal} {\bibinfo  {journal} {ACS Photonics}\ }\textbf
  {\bibinfo {volume} {9}},\ \bibinfo {pages} {2398} (\bibinfo {year}
  {2022})}\BibitemShut {NoStop}%
\bibitem [{\citenamefont {Zotev}\ \emph {et~al.}(2023)\citenamefont {Zotev},
  \citenamefont {Wang}, \citenamefont {Andres-Penares}, \citenamefont
  {Severs-Millard}, \citenamefont {Randerson}, \citenamefont {Hu},
  \citenamefont {Sortino}, \citenamefont {Louca}, \citenamefont
  {Brotons-Gisbert}, \citenamefont {Huq} \emph {et~al.}}]{Zotev2023}%
  \BibitemOpen
  \bibfield  {author} {\bibinfo {author} {\bibfnamefont {P.~G.}\ \bibnamefont
  {Zotev}}, \bibinfo {author} {\bibfnamefont {Y.}~\bibnamefont {Wang}},
  \bibinfo {author} {\bibfnamefont {D.}~\bibnamefont {Andres-Penares}},
  \bibinfo {author} {\bibfnamefont {T.}~\bibnamefont {Severs-Millard}},
  \bibinfo {author} {\bibfnamefont {S.}~\bibnamefont {Randerson}}, \bibinfo
  {author} {\bibfnamefont {X.}~\bibnamefont {Hu}}, \bibinfo {author}
  {\bibfnamefont {L.}~\bibnamefont {Sortino}}, \bibinfo {author} {\bibfnamefont
  {C.}~\bibnamefont {Louca}}, \bibinfo {author} {\bibfnamefont
  {M.}~\bibnamefont {Brotons-Gisbert}}, \bibinfo {author} {\bibfnamefont
  {T.}~\bibnamefont {Huq}}, \emph {et~al.},\ }\bibfield  {title} {\bibinfo
  {title} {Van der waals materials for applications in nanophotonics},\
  }\href@noop {} {\bibfield  {journal} {\bibinfo  {journal} {Laser \& Photonics
  Reviews}\ }\textbf {\bibinfo {volume} {17}},\ \bibinfo {pages} {2200957}
  (\bibinfo {year} {2023})}\BibitemShut {NoStop}%
\bibitem [{\citenamefont {Mounet}\ \emph {et~al.}(2018)\citenamefont {Mounet},
  \citenamefont {Gibertini}, \citenamefont {Schwaller}, \citenamefont {Campi},
  \citenamefont {Merkys}, \citenamefont {Marrazzo}, \citenamefont {Sohier},
  \citenamefont {Castelli}, \citenamefont {Cepellotti}, \citenamefont {Pizzi},\
  and\ \citenamefont {Marzari}}]{Mounet2018}%
  \BibitemOpen
  \bibfield  {author} {\bibinfo {author} {\bibfnamefont {N.}~\bibnamefont
  {Mounet}}, \bibinfo {author} {\bibfnamefont {M.}~\bibnamefont {Gibertini}},
  \bibinfo {author} {\bibfnamefont {P.}~\bibnamefont {Schwaller}}, \bibinfo
  {author} {\bibfnamefont {D.}~\bibnamefont {Campi}}, \bibinfo {author}
  {\bibfnamefont {A.}~\bibnamefont {Merkys}}, \bibinfo {author} {\bibfnamefont
  {A.}~\bibnamefont {Marrazzo}}, \bibinfo {author} {\bibfnamefont
  {T.}~\bibnamefont {Sohier}}, \bibinfo {author} {\bibfnamefont {I.~E.}\
  \bibnamefont {Castelli}}, \bibinfo {author} {\bibfnamefont {A.}~\bibnamefont
  {Cepellotti}}, \bibinfo {author} {\bibfnamefont {G.}~\bibnamefont {Pizzi}},\
  and\ \bibinfo {author} {\bibfnamefont {N.}~\bibnamefont {Marzari}},\
  }\bibfield  {title} {\bibinfo {title} {{Two-dimensional materials from
  high-throughput computational exfoliation of experimentally known
  compounds}},\ }\href {https://doi.org/10.1038/s41565-017-0035-5} {\bibfield
  {journal} {\bibinfo  {journal} {Nat. Nanotechnol.}\ }\textbf {\bibinfo
  {volume} {13}},\ \bibinfo {pages} {246} (\bibinfo {year} {2018})}\BibitemShut
  {NoStop}%
\bibitem [{\citenamefont {Haastrup}\ \emph {et~al.}(2018)\citenamefont
  {Haastrup}, \citenamefont {Strange}, \citenamefont {Pandey}, \citenamefont
  {Deilmann}, \citenamefont {Schmidt}, \citenamefont {Hinsche}, \citenamefont
  {Gjerding}, \citenamefont {Torelli}, \citenamefont {Larsen}, \citenamefont
  {Riis-Jensen} \emph {et~al.}}]{Haastrup2018}%
  \BibitemOpen
  \bibfield  {author} {\bibinfo {author} {\bibfnamefont {S.}~\bibnamefont
  {Haastrup}}, \bibinfo {author} {\bibfnamefont {M.}~\bibnamefont {Strange}},
  \bibinfo {author} {\bibfnamefont {M.}~\bibnamefont {Pandey}}, \bibinfo
  {author} {\bibfnamefont {T.}~\bibnamefont {Deilmann}}, \bibinfo {author}
  {\bibfnamefont {P.~S.}\ \bibnamefont {Schmidt}}, \bibinfo {author}
  {\bibfnamefont {N.~F.}\ \bibnamefont {Hinsche}}, \bibinfo {author}
  {\bibfnamefont {M.~N.}\ \bibnamefont {Gjerding}}, \bibinfo {author}
  {\bibfnamefont {D.}~\bibnamefont {Torelli}}, \bibinfo {author} {\bibfnamefont
  {P.~M.}\ \bibnamefont {Larsen}}, \bibinfo {author} {\bibfnamefont {A.~C.}\
  \bibnamefont {Riis-Jensen}}, \emph {et~al.},\ }\bibfield  {title} {\bibinfo
  {title} {The computational 2D materials database: high-throughput modeling
  and discovery of atomically thin crystals},\ }\href@noop {} {\bibfield
  {journal} {\bibinfo  {journal} {2D Materials}\ }\textbf {\bibinfo {volume}
  {5}},\ \bibinfo {pages} {042002} (\bibinfo {year} {2018})}\BibitemShut
  {NoStop}%
\bibitem [{\citenamefont {N{\o}rskov}\ \emph {et~al.}(2009)\citenamefont
  {N{\o}rskov}, \citenamefont {Bligaard}, \citenamefont {Rossmeisl},\ and\
  \citenamefont {Christensen}}]{Nørskov2009}%
  \BibitemOpen
  \bibfield  {author} {\bibinfo {author} {\bibfnamefont {J.~K.}\ \bibnamefont
  {N{\o}rskov}}, \bibinfo {author} {\bibfnamefont {T.}~\bibnamefont
  {Bligaard}}, \bibinfo {author} {\bibfnamefont {J.}~\bibnamefont
  {Rossmeisl}},\ and\ \bibinfo {author} {\bibfnamefont {C.~H.}\ \bibnamefont
  {Christensen}},\ }\bibfield  {title} {\bibinfo {title} {{Towards the
  computational design of solid catalysts}},\ }\href
  {https://doi.org/10.1038/nchem.121} {\bibfield  {journal} {\bibinfo
  {journal} {Nat. Chem.}\ }\textbf {\bibinfo {volume} {1}},\ \bibinfo {pages}
  {37} (\bibinfo {year} {2009})}\BibitemShut {NoStop}%
\bibitem [{\citenamefont {Xiao}\ \emph {et~al.}(2019)\citenamefont {Xiao},
  \citenamefont {Miara}, \citenamefont {Wang},\ and\ \citenamefont
  {Ceder}}]{Xiao2019}%
  \BibitemOpen
  \bibfield  {author} {\bibinfo {author} {\bibfnamefont {Y.}~\bibnamefont
  {Xiao}}, \bibinfo {author} {\bibfnamefont {L.~J.}\ \bibnamefont {Miara}},
  \bibinfo {author} {\bibfnamefont {Y.}~\bibnamefont {Wang}},\ and\ \bibinfo
  {author} {\bibfnamefont {G.}~\bibnamefont {Ceder}},\ }\bibfield  {title}
  {\bibinfo {title} {{Computational Screening of Cathode Coatings for
  Solid-State Batteries}},\ }\href
  {https://doi.org/10.1016/j.joule.2019.02.006} {\bibfield  {journal} {\bibinfo
   {journal} {Joule}\ }\textbf {\bibinfo {volume} {3}},\ \bibinfo {pages}
  {1252} (\bibinfo {year} {2019})}\BibitemShut {NoStop}%
\bibitem [{\citenamefont {Yu}\ and\ \citenamefont {Zunger}(2012)}]{Yu2012}%
  \BibitemOpen
  \bibfield  {author} {\bibinfo {author} {\bibfnamefont {L.}~\bibnamefont
  {Yu}}\ and\ \bibinfo {author} {\bibfnamefont {A.}~\bibnamefont {Zunger}},\
  }\bibfield  {title} {\bibinfo {title} {{Identification of potential
  photovoltaic absorbers based on first-principles spectroscopic screening of
  materials}},\ }\href {https://doi.org/10.1103/PhysRevLett.108.068701}
  {\bibfield  {journal} {\bibinfo  {journal} {Phys. Rev. Lett.}\ }\textbf
  {\bibinfo {volume} {108}},\ \bibinfo {pages} {068701} (\bibinfo {year}
  {2012})}\BibitemShut {NoStop}%
\bibitem [{\citenamefont {Kuhar}\ \emph {et~al.}(2017)\citenamefont {Kuhar},
  \citenamefont {Crovetto}, \citenamefont {Pandey}, \citenamefont {Thygesen},
  \citenamefont {Seger}, \citenamefont {Vesborg}, \citenamefont {Hansen},
  \citenamefont {Chorkendorff},\ and\ \citenamefont {Jacobsen}}]{Kuhar2017}%
  \BibitemOpen
  \bibfield  {author} {\bibinfo {author} {\bibfnamefont {K.}~\bibnamefont
  {Kuhar}}, \bibinfo {author} {\bibfnamefont {A.}~\bibnamefont {Crovetto}},
  \bibinfo {author} {\bibfnamefont {M.}~\bibnamefont {Pandey}}, \bibinfo
  {author} {\bibfnamefont {K.~S.}\ \bibnamefont {Thygesen}}, \bibinfo {author}
  {\bibfnamefont {B.}~\bibnamefont {Seger}}, \bibinfo {author} {\bibfnamefont
  {P.~C.~K.}\ \bibnamefont {Vesborg}}, \bibinfo {author} {\bibfnamefont
  {O.}~\bibnamefont {Hansen}}, \bibinfo {author} {\bibfnamefont
  {I.}~\bibnamefont {Chorkendorff}},\ and\ \bibinfo {author} {\bibfnamefont
  {K.~W.}\ \bibnamefont {Jacobsen}},\ }\bibfield  {title} {\bibinfo {title}
  {{Sulfide perovskites for solar energy conversion applications: computational
  screening and synthesis of the selected compound LaYS3}},\ }\href
  {https://doi.org/10.1039/C7EE02702H} {\bibfield  {journal} {\bibinfo
  {journal} {Energy Environ. Sci.}\ }\textbf {\bibinfo {volume} {10}},\
  \bibinfo {pages} {2579} (\bibinfo {year} {2017})}\BibitemShut {NoStop}%
\bibitem [{\citenamefont {Naccarato}\ \emph {et~al.}(2019)\citenamefont
  {Naccarato}, \citenamefont {Ricci}, \citenamefont {Suntivich}, \citenamefont
  {Hautier}, \citenamefont {Wirtz},\ and\ \citenamefont
  {Rignanese}}]{Naccarato2019}%
  \BibitemOpen
  \bibfield  {author} {\bibinfo {author} {\bibfnamefont {F.}~\bibnamefont
  {Naccarato}}, \bibinfo {author} {\bibfnamefont {F.}~\bibnamefont {Ricci}},
  \bibinfo {author} {\bibfnamefont {J.}~\bibnamefont {Suntivich}}, \bibinfo
  {author} {\bibfnamefont {G.}~\bibnamefont {Hautier}}, \bibinfo {author}
  {\bibfnamefont {L.}~\bibnamefont {Wirtz}},\ and\ \bibinfo {author}
  {\bibfnamefont {G.-M.}\ \bibnamefont {Rignanese}},\ }\bibfield  {title}
  {\bibinfo {title} {{Searching for materials with high refractive index and
  wide band gap: A first-principles high-throughput study}},\ }\href
  {https://doi.org/10.1103/PhysRevMaterials.3.044602} {\bibfield  {journal}
  {\bibinfo  {journal} {Phys. Rev. Mater.}\ }\textbf {\bibinfo {volume} {3}},\
  \bibinfo {pages} {044602} (\bibinfo {year} {2019})}\BibitemShut {NoStop}%
\bibitem [{\citenamefont {Svendsen}\ \emph {et~al.}(2022)\citenamefont
  {Svendsen}, \citenamefont {Sugimoto}, \citenamefont {Assadillayev},
  \citenamefont {Shima}, \citenamefont {Fujii}, \citenamefont {Thygesen},\ and\
  \citenamefont {Raza}}]{Svendsen2022}%
  \BibitemOpen
  \bibfield  {author} {\bibinfo {author} {\bibfnamefont {M.~K.}\ \bibnamefont
  {Svendsen}}, \bibinfo {author} {\bibfnamefont {H.}~\bibnamefont {Sugimoto}},
  \bibinfo {author} {\bibfnamefont {A.}~\bibnamefont {Assadillayev}}, \bibinfo
  {author} {\bibfnamefont {D.}~\bibnamefont {Shima}}, \bibinfo {author}
  {\bibfnamefont {M.}~\bibnamefont {Fujii}}, \bibinfo {author} {\bibfnamefont
  {K.~S.}\ \bibnamefont {Thygesen}},\ and\ \bibinfo {author} {\bibfnamefont
  {S.}~\bibnamefont {Raza}},\ }\bibfield  {title} {\bibinfo {title}
  {Computational discovery and experimental demonstration of boron phosphide
  ultraviolet nanoresonators},\ }\href@noop {} {\bibfield  {journal} {\bibinfo
  {journal} {Advanced Optical Materials}\ }\textbf {\bibinfo {volume} {10}},\
  \bibinfo {pages} {2200422} (\bibinfo {year} {2022})}\BibitemShut {NoStop}%
\bibitem [{\citenamefont {Albrecht}\ \emph {et~al.}(1998)\citenamefont
  {Albrecht}, \citenamefont {Reining}, \citenamefont {{Del Sole}},\ and\
  \citenamefont {Onida}}]{Albrecht1998}%
  \BibitemOpen
  \bibfield  {author} {\bibinfo {author} {\bibfnamefont {S.}~\bibnamefont
  {Albrecht}}, \bibinfo {author} {\bibfnamefont {L.}~\bibnamefont {Reining}},
  \bibinfo {author} {\bibfnamefont {R.}~\bibnamefont {{Del Sole}}},\ and\
  \bibinfo {author} {\bibfnamefont {G.}~\bibnamefont {Onida}},\ }\bibfield
  {title} {\bibinfo {title} {{Ab Initio Calculation of Excitonic Effects in the
  Optical Spectra of Semiconductors}},\ }\href
  {https://doi.org/10.1103/PhysRevLett.80.4510} {\bibfield  {journal} {\bibinfo
   {journal} {Phys. Rev. Lett.}\ }\textbf {\bibinfo {volume} {80}},\ \bibinfo
  {pages} {4510} (\bibinfo {year} {1998})}\BibitemShut {NoStop}%
\bibitem [{\citenamefont {S{\o}ndersted}\ \emph
  {et~al.}(2024{\natexlab{a}})\citenamefont {S{\o}ndersted}, \citenamefont
  {Kuisma}, \citenamefont {Svaneborg}, \citenamefont {Svendsen},\ and\
  \citenamefont {Thygesen}}]{Søndersted2024}%
  \BibitemOpen
  \bibfield  {author} {\bibinfo {author} {\bibfnamefont {A.~H.}\ \bibnamefont
  {S{\o}ndersted}}, \bibinfo {author} {\bibfnamefont {M.}~\bibnamefont
  {Kuisma}}, \bibinfo {author} {\bibfnamefont {J.~K.}\ \bibnamefont
  {Svaneborg}}, \bibinfo {author} {\bibfnamefont {M.~K.}\ \bibnamefont
  {Svendsen}},\ and\ \bibinfo {author} {\bibfnamefont {K.~S.}\ \bibnamefont
  {Thygesen}},\ }\bibfield  {title} {\bibinfo {title} {{Improved Dielectric
  Response of Solids: Combining the Bethe-Salpeter Equation with the Random
  Phase Approximation}},\ }\href
  {https://doi.org/10.1103/PhysRevLett.133.026403} {\bibfield  {journal}
  {\bibinfo  {journal} {Phys. Rev. Lett.}\ }\textbf {\bibinfo {volume} {133}},\
  \bibinfo {pages} {026403} (\bibinfo {year} {2024}{\natexlab{a}})}\BibitemShut
  {NoStop}%
\bibitem [{\citenamefont {Doiron}\ \emph {et~al.}(2022)\citenamefont {Doiron},
  \citenamefont {Khurgin},\ and\ \citenamefont {Naik}}]{Doiron2022}%
  \BibitemOpen
  \bibfield  {author} {\bibinfo {author} {\bibfnamefont {C.~F.}\ \bibnamefont
  {Doiron}}, \bibinfo {author} {\bibfnamefont {J.~B.}\ \bibnamefont
  {Khurgin}},\ and\ \bibinfo {author} {\bibfnamefont {G.~V.}\ \bibnamefont
  {Naik}},\ }\bibfield  {title} {\bibinfo {title} {Super-mossian dielectrics
  for nanophotonics},\ }\href@noop {} {\bibfield  {journal} {\bibinfo
  {journal} {Advanced Optical Materials}\ }\textbf {\bibinfo {volume} {10}},\
  \bibinfo {pages} {2201084} (\bibinfo {year} {2022})}\BibitemShut {NoStop}%
\bibitem [{\citenamefont {Lee}\ \emph {et~al.}(2019)\citenamefont {Lee},
  \citenamefont {Jeong}, \citenamefont {Jung},\ and\ \citenamefont
  {Yee}}]{Lee2019}%
  \BibitemOpen
  \bibfield  {author} {\bibinfo {author} {\bibfnamefont {S.-Y.}\ \bibnamefont
  {Lee}}, \bibinfo {author} {\bibfnamefont {T.-Y.}\ \bibnamefont {Jeong}},
  \bibinfo {author} {\bibfnamefont {S.}~\bibnamefont {Jung}},\ and\ \bibinfo
  {author} {\bibfnamefont {K.-J.}\ \bibnamefont {Yee}},\ }\bibfield  {title}
  {\bibinfo {title} {Refractive index dispersion of hexagonal boron nitride in
  the visible and near-infrared},\ }\href@noop {} {\bibfield  {journal}
  {\bibinfo  {journal} {Phys. Status Solidi B}\ }\textbf {\bibinfo {volume}
  {256}},\ \bibinfo {pages} {1800417} (\bibinfo {year} {2019})}\BibitemShut
  {NoStop}%
\bibitem [{\citenamefont {Vyshnevyy}\ \emph {et~al.}(2023)\citenamefont
  {Vyshnevyy}, \citenamefont {Ermolaev}, \citenamefont {Grudinin},
  \citenamefont {Voronin}, \citenamefont {Kharichkin}, \citenamefont {Mazitov},
  \citenamefont {Kruglov}, \citenamefont {Yakubovsky}, \citenamefont {Mishra},
  \citenamefont {Kirtaev} \emph {et~al.}}]{Vyshnevyy2023}%
  \BibitemOpen
  \bibfield  {author} {\bibinfo {author} {\bibfnamefont {A.~A.}\ \bibnamefont
  {Vyshnevyy}}, \bibinfo {author} {\bibfnamefont {G.~A.}\ \bibnamefont
  {Ermolaev}}, \bibinfo {author} {\bibfnamefont {D.~V.}\ \bibnamefont
  {Grudinin}}, \bibinfo {author} {\bibfnamefont {K.~V.}\ \bibnamefont
  {Voronin}}, \bibinfo {author} {\bibfnamefont {I.}~\bibnamefont {Kharichkin}},
  \bibinfo {author} {\bibfnamefont {A.}~\bibnamefont {Mazitov}}, \bibinfo
  {author} {\bibfnamefont {I.~A.}\ \bibnamefont {Kruglov}}, \bibinfo {author}
  {\bibfnamefont {D.~I.}\ \bibnamefont {Yakubovsky}}, \bibinfo {author}
  {\bibfnamefont {P.}~\bibnamefont {Mishra}}, \bibinfo {author} {\bibfnamefont
  {R.~V.}\ \bibnamefont {Kirtaev}}, \emph {et~al.},\ }\bibfield  {title}
  {\bibinfo {title} {Van der waals materials for overcoming fundamental
  limitations in photonic integrated circuitry},\ }\href@noop {} {\bibfield
  {journal} {\bibinfo  {journal} {Nano Lett.}\ }\textbf {\bibinfo {volume}
  {23}},\ \bibinfo {pages} {8057} (\bibinfo {year} {2023})}\BibitemShut
  {NoStop}%
\bibitem [{\citenamefont {Polyanskiy}(2024)}]{Polyanskiy2024}%
  \BibitemOpen
  \bibfield  {author} {\bibinfo {author} {\bibfnamefont {M.~N.}\ \bibnamefont
  {Polyanskiy}},\ }\bibfield  {title} {\bibinfo {title} {Refractiveindex.info
  database of optical constants},\ }\href@noop {} {\bibfield  {journal}
  {\bibinfo  {journal} {Sci. Data}\ }\textbf {\bibinfo {volume} {11}},\
  \bibinfo {pages} {94} (\bibinfo {year} {2024})}\BibitemShut {NoStop}%
\bibitem [{\citenamefont {Kirklin}\ \emph {et~al.}(2015)\citenamefont
  {Kirklin}, \citenamefont {Saal}, \citenamefont {Meredig}, \citenamefont
  {Thompson}, \citenamefont {Doak}, \citenamefont {Aykol}, \citenamefont
  {R{\"u}hl},\ and\ \citenamefont {Wolverton}}]{kirklin2015open}%
  \BibitemOpen
  \bibfield  {author} {\bibinfo {author} {\bibfnamefont {S.}~\bibnamefont
  {Kirklin}}, \bibinfo {author} {\bibfnamefont {J.~E.}\ \bibnamefont {Saal}},
  \bibinfo {author} {\bibfnamefont {B.}~\bibnamefont {Meredig}}, \bibinfo
  {author} {\bibfnamefont {A.}~\bibnamefont {Thompson}}, \bibinfo {author}
  {\bibfnamefont {J.~W.}\ \bibnamefont {Doak}}, \bibinfo {author}
  {\bibfnamefont {M.}~\bibnamefont {Aykol}}, \bibinfo {author} {\bibfnamefont
  {S.}~\bibnamefont {R{\"u}hl}},\ and\ \bibinfo {author} {\bibfnamefont
  {C.}~\bibnamefont {Wolverton}},\ }\bibfield  {title} {\bibinfo {title} {{The
  Open Quantum Materials Database (OQMD): assessing the accuracy of DFT
  formation energies}},\ }\href@noop {} {\bibfield  {journal} {\bibinfo
  {journal} {NPJ Comput. Mater.}\ }\textbf {\bibinfo {volume} {1}},\ \bibinfo
  {pages} {15010} (\bibinfo {year} {2015})}\BibitemShut {NoStop}%
\bibitem [{\citenamefont {Perdew}\ \emph {et~al.}(1996)\citenamefont {Perdew},
  \citenamefont {Burke},\ and\ \citenamefont {Ernzerhof}}]{PBE}%
  \BibitemOpen
  \bibfield  {author} {\bibinfo {author} {\bibfnamefont {J.~P.}\ \bibnamefont
  {Perdew}}, \bibinfo {author} {\bibfnamefont {K.}~\bibnamefont {Burke}},\ and\
  \bibinfo {author} {\bibfnamefont {M.}~\bibnamefont {Ernzerhof}},\ }\bibfield
  {title} {\bibinfo {title} {Generalized gradient approximation made simple},\
  }\href@noop {} {\bibfield  {journal} {\bibinfo  {journal} {Phys. Rev. Lett.}\
  }\textbf {\bibinfo {volume} {77}},\ \bibinfo {pages} {3865} (\bibinfo {year}
  {1996})}\BibitemShut {NoStop}%
\bibitem [{\citenamefont {Grimme}\ \emph {et~al.}(2010)\citenamefont {Grimme},
  \citenamefont {Antony}, \citenamefont {Ehrlich},\ and\ \citenamefont
  {Krieg}}]{DFTD3}%
  \BibitemOpen
  \bibfield  {author} {\bibinfo {author} {\bibfnamefont {S.}~\bibnamefont
  {Grimme}}, \bibinfo {author} {\bibfnamefont {J.}~\bibnamefont {Antony}},
  \bibinfo {author} {\bibfnamefont {S.}~\bibnamefont {Ehrlich}},\ and\ \bibinfo
  {author} {\bibfnamefont {H.}~\bibnamefont {Krieg}},\ }\bibfield  {title}
  {\bibinfo {title} {{A consistent and accurate ab initio parametrization of
  density functional dispersion correction (DFT-D) for the 94 elements H-Pu}},\
  }\href@noop {} {\bibfield  {journal} {\bibinfo  {journal} {J. Chem. Phys.}\
  }\textbf {\bibinfo {volume} {132}},\ \bibinfo {pages} {154104} (\bibinfo
  {year} {2010})}\BibitemShut {NoStop}%
\bibitem [{\citenamefont {CAMD}(2025)}]{CRYSPdatabase}%
  \BibitemOpen
  \bibfield  {author} {\bibinfo {author} {\bibnamefont {CAMD}},\ }\href@noop {}
  {\bibinfo {title} {A computational database of crystal properties}} (\bibinfo
  {year} {2025}),\ \bibinfo {note} {available at
  \url{https://crysp.fysik.dtu.dk/}}\BibitemShut {NoStop}%
\bibitem [{\citenamefont {Perdew}(1985)}]{perdew1985density}%
  \BibitemOpen
  \bibfield  {author} {\bibinfo {author} {\bibfnamefont {J.~P.}\ \bibnamefont
  {Perdew}},\ }\bibfield  {title} {\bibinfo {title} {Density functional theory
  and the band gap problem},\ }\href@noop {} {\bibfield  {journal} {\bibinfo
  {journal} {Int. J. Quantum Chem.}\ }\textbf {\bibinfo {volume} {28}},\
  \bibinfo {pages} {497} (\bibinfo {year} {1985})}\BibitemShut {NoStop}%
\bibitem [{\citenamefont {Chae}\ \emph {et~al.}(2016)\citenamefont {Chae},
  \citenamefont {Jin}, \citenamefont {Kim}, \citenamefont {Chung},
  \citenamefont {Na}, \citenamefont {Nam}, \citenamefont {Kim}, \citenamefont
  {Perello}, \citenamefont {Jeong}, \citenamefont {Ly} \emph
  {et~al.}}]{Chae2016}%
  \BibitemOpen
  \bibfield  {author} {\bibinfo {author} {\bibfnamefont {S.~H.}\ \bibnamefont
  {Chae}}, \bibinfo {author} {\bibfnamefont {Y.}~\bibnamefont {Jin}}, \bibinfo
  {author} {\bibfnamefont {T.~S.}\ \bibnamefont {Kim}}, \bibinfo {author}
  {\bibfnamefont {D.~S.}\ \bibnamefont {Chung}}, \bibinfo {author}
  {\bibfnamefont {H.}~\bibnamefont {Na}}, \bibinfo {author} {\bibfnamefont
  {H.}~\bibnamefont {Nam}}, \bibinfo {author} {\bibfnamefont {H.}~\bibnamefont
  {Kim}}, \bibinfo {author} {\bibfnamefont {D.~J.}\ \bibnamefont {Perello}},
  \bibinfo {author} {\bibfnamefont {H.~Y.}\ \bibnamefont {Jeong}}, \bibinfo
  {author} {\bibfnamefont {T.~H.}\ \bibnamefont {Ly}}, \emph {et~al.},\
  }\bibfield  {title} {\bibinfo {title} {Oxidation effect in octahedral hafnium
  disulfide thin film},\ }\href@noop {} {\bibfield  {journal} {\bibinfo
  {journal} {ACS Nano}\ }\textbf {\bibinfo {volume} {10}},\ \bibinfo {pages}
  {1309} (\bibinfo {year} {2016})}\BibitemShut {NoStop}%
\bibitem [{\citenamefont {Mirabelli}\ \emph {et~al.}(2016)\citenamefont
  {Mirabelli}, \citenamefont {McGeough}, \citenamefont {Schmidt}, \citenamefont
  {McCarthy}, \citenamefont {Monaghan}, \citenamefont {Povey}, \citenamefont
  {McCarthy}, \citenamefont {Gity}, \citenamefont {Nagle}, \citenamefont
  {Hughes} \emph {et~al.}}]{Mirabelli2016}%
  \BibitemOpen
  \bibfield  {author} {\bibinfo {author} {\bibfnamefont {G.}~\bibnamefont
  {Mirabelli}}, \bibinfo {author} {\bibfnamefont {C.}~\bibnamefont {McGeough}},
  \bibinfo {author} {\bibfnamefont {M.}~\bibnamefont {Schmidt}}, \bibinfo
  {author} {\bibfnamefont {E.~K.}\ \bibnamefont {McCarthy}}, \bibinfo {author}
  {\bibfnamefont {S.}~\bibnamefont {Monaghan}}, \bibinfo {author}
  {\bibfnamefont {I.~M.}\ \bibnamefont {Povey}}, \bibinfo {author}
  {\bibfnamefont {M.}~\bibnamefont {McCarthy}}, \bibinfo {author}
  {\bibfnamefont {F.}~\bibnamefont {Gity}}, \bibinfo {author} {\bibfnamefont
  {R.}~\bibnamefont {Nagle}}, \bibinfo {author} {\bibfnamefont
  {G.}~\bibnamefont {Hughes}}, \emph {et~al.},\ }\bibfield  {title} {\bibinfo
  {title} {{Air sensitivity of MoS$_2$, MoSe$_2$, MoTe$_2$, HfS$_2$, and
  HfSe$_2$}},\ }\href@noop {} {\bibfield  {journal} {\bibinfo  {journal} {J.
  Appl. Phys.}\ }\textbf {\bibinfo {volume} {120}} (\bibinfo {year}
  {2016})}\BibitemShut {NoStop}%
\bibitem [{\citenamefont {Lai}\ \emph {et~al.}(2018)\citenamefont {Lai},
  \citenamefont {Byeon}, \citenamefont {Jang}, \citenamefont {Lee},
  \citenamefont {Lee}, \citenamefont {Park}, \citenamefont {Kim},\ and\
  \citenamefont {Lee}}]{Lai2018}%
  \BibitemOpen
  \bibfield  {author} {\bibinfo {author} {\bibfnamefont {S.}~\bibnamefont
  {Lai}}, \bibinfo {author} {\bibfnamefont {S.}~\bibnamefont {Byeon}}, \bibinfo
  {author} {\bibfnamefont {S.~K.}\ \bibnamefont {Jang}}, \bibinfo {author}
  {\bibfnamefont {J.}~\bibnamefont {Lee}}, \bibinfo {author} {\bibfnamefont
  {B.~H.}\ \bibnamefont {Lee}}, \bibinfo {author} {\bibfnamefont {J.-H.}\
  \bibnamefont {Park}}, \bibinfo {author} {\bibfnamefont {Y.-H.}\ \bibnamefont
  {Kim}},\ and\ \bibinfo {author} {\bibfnamefont {S.}~\bibnamefont {Lee}},\
  }\bibfield  {title} {\bibinfo {title} {{HfO$_2$/HfS$_2$ hybrid
  heterostructure fabricated via controllable chemical conversion of
  two-dimensional HfS$_2$}},\ }\href@noop {} {\bibfield  {journal} {\bibinfo
  {journal} {Nanoscale}\ }\textbf {\bibinfo {volume} {10}},\ \bibinfo {pages}
  {18758} (\bibinfo {year} {2018})}\BibitemShut {NoStop}%
\bibitem [{\citenamefont {Jin}\ \emph {et~al.}(2021)\citenamefont {Jin},
  \citenamefont {Zheng}, \citenamefont {Gao}, \citenamefont {Wang},
  \citenamefont {Li}, \citenamefont {Chen}, \citenamefont {Pan}, \citenamefont
  {Lin},\ and\ \citenamefont {Chen}}]{Jin2021}%
  \BibitemOpen
  \bibfield  {author} {\bibinfo {author} {\bibfnamefont {T.}~\bibnamefont
  {Jin}}, \bibinfo {author} {\bibfnamefont {Y.}~\bibnamefont {Zheng}}, \bibinfo
  {author} {\bibfnamefont {J.}~\bibnamefont {Gao}}, \bibinfo {author}
  {\bibfnamefont {Y.}~\bibnamefont {Wang}}, \bibinfo {author} {\bibfnamefont
  {E.}~\bibnamefont {Li}}, \bibinfo {author} {\bibfnamefont {H.}~\bibnamefont
  {Chen}}, \bibinfo {author} {\bibfnamefont {X.}~\bibnamefont {Pan}}, \bibinfo
  {author} {\bibfnamefont {M.}~\bibnamefont {Lin}},\ and\ \bibinfo {author}
  {\bibfnamefont {W.}~\bibnamefont {Chen}},\ }\bibfield  {title} {\bibinfo
  {title} {{Controlling native oxidation of HfS$_2$ for 2D materials based
  flash memory and artificial synapse}},\ }\href@noop {} {\bibfield  {journal}
  {\bibinfo  {journal} {ACS Appl. Mater. Interfaces}\ }\textbf {\bibinfo
  {volume} {13}},\ \bibinfo {pages} {10639} (\bibinfo {year}
  {2021})}\BibitemShut {NoStop}%
\bibitem [{\citenamefont {Danielsen}\ \emph {et~al.}(2021)\citenamefont
  {Danielsen}, \citenamefont {Lyksborg-Andersen}, \citenamefont {Nielsen},
  \citenamefont {Jessen}, \citenamefont {Booth}, \citenamefont {Doan},
  \citenamefont {Zhou}, \citenamefont {B{\o}ggild},\ and\ \citenamefont
  {Gammelgaard}}]{Danielsen2021}%
  \BibitemOpen
  \bibfield  {author} {\bibinfo {author} {\bibfnamefont {D.~R.}\ \bibnamefont
  {Danielsen}}, \bibinfo {author} {\bibfnamefont {A.}~\bibnamefont
  {Lyksborg-Andersen}}, \bibinfo {author} {\bibfnamefont {K.~E.~S.}\
  \bibnamefont {Nielsen}}, \bibinfo {author} {\bibfnamefont {B.~S.}\
  \bibnamefont {Jessen}}, \bibinfo {author} {\bibfnamefont {T.~J.}\
  \bibnamefont {Booth}}, \bibinfo {author} {\bibfnamefont {M.-H.}\ \bibnamefont
  {Doan}}, \bibinfo {author} {\bibfnamefont {Y.}~\bibnamefont {Zhou}}, \bibinfo
  {author} {\bibfnamefont {P.}~\bibnamefont {B{\o}ggild}},\ and\ \bibinfo
  {author} {\bibfnamefont {L.}~\bibnamefont {Gammelgaard}},\ }\bibfield
  {title} {\bibinfo {title} {{Super-Resolution Nanolithography of
  Two-Dimensional Materials by Anisotropic Etching}},\ }\href
  {https://doi.org/10.1021/acsami.1c09923} {\bibfield  {journal} {\bibinfo
  {journal} {ACS Appl. Mater. Interfaces}\ }\textbf {\bibinfo {volume} {13}},\
  \bibinfo {pages} {41886} (\bibinfo {year} {2021})}\BibitemShut {NoStop}%
\bibitem [{Sar(2024)}]{Sarbajna2024}%
  \BibitemOpen
  \bibfield  {title} {\bibinfo {title} {{Encapsulated Void Resonators in Van
  der Waals Heterostructures}},\ }\href
  {https://doi.org/10.1002/lpor.202401215} {\bibfield  {journal} {\bibinfo
  {journal} {Laser Photon. Rev.}\ ,\ \bibinfo {pages} {2401215}} (\bibinfo
  {year} {2024})}\BibitemShut {NoStop}%
\bibitem [{\citenamefont {Allen~Jr}(2014)}]{AllenThesis2014}%
  \BibitemOpen
  \bibfield  {author} {\bibinfo {author} {\bibfnamefont {K.~W.}\ \bibnamefont
  {Allen~Jr}},\ }\bibfield  {title} {\bibinfo {title} {Waveguide,
  photodetector, and imaging applications of microspherical photonics},\
  }\href@noop {} {\  (\bibinfo {year} {2014})}\BibitemShut {NoStop}%
\bibitem [{\citenamefont {Darafsheh}(2021)}]{Darafsheh2021}%
  \BibitemOpen
  \bibfield  {author} {\bibinfo {author} {\bibfnamefont {A.}~\bibnamefont
  {Darafsheh}},\ }\bibfield  {title} {\bibinfo {title} {Photonic nanojets and
  their applications},\ }\href@noop {} {\bibfield  {journal} {\bibinfo
  {journal} {JPhys: Photonics}\ }\textbf {\bibinfo {volume} {3}},\
  \bibinfo {pages} {022001} (\bibinfo {year} {2021})}\BibitemShut {NoStop}%
\bibitem [{\citenamefont {Surdo}\ \emph {et~al.}(2021)\citenamefont {Surdo},
  \citenamefont {Duocastella},\ and\ \citenamefont {Diaspro}}]{Surdo2021}%
  \BibitemOpen
  \bibfield  {author} {\bibinfo {author} {\bibfnamefont {S.}~\bibnamefont
  {Surdo}}, \bibinfo {author} {\bibfnamefont {M.}~\bibnamefont {Duocastella}},\
  and\ \bibinfo {author} {\bibfnamefont {A.}~\bibnamefont {Diaspro}},\
  }\bibfield  {title} {\bibinfo {title} {Nanopatterning with photonic nanojets:
  Review and perspectives in biomedical research},\ }\bibfield  {journal}
  {\bibinfo  {journal} {Micromachines}\ }\textbf {\bibinfo {volume} {12}},\
  \href {https://doi.org/10.3390/mi12030256} {10.3390/mi12030256} (\bibinfo
  {year} {2021})\BibitemShut {NoStop}%
\bibitem [{\citenamefont {Larsen}\ \emph {et~al.}(2017)\citenamefont {Larsen},
  \citenamefont {Mortensen}, \citenamefont {Blomqvist}, \citenamefont
  {Castelli}, \citenamefont {Christensen}, \citenamefont {Du{\l}ak},
  \citenamefont {Friis}, \citenamefont {Groves}, \citenamefont {Hammer},
  \citenamefont {Hargus} \emph {et~al.}}]{Larsen2017}%
  \BibitemOpen
  \bibfield  {author} {\bibinfo {author} {\bibfnamefont {A.~H.}\ \bibnamefont
  {Larsen}}, \bibinfo {author} {\bibfnamefont {J.~J.}\ \bibnamefont
  {Mortensen}}, \bibinfo {author} {\bibfnamefont {J.}~\bibnamefont
  {Blomqvist}}, \bibinfo {author} {\bibfnamefont {I.~E.}\ \bibnamefont
  {Castelli}}, \bibinfo {author} {\bibfnamefont {R.}~\bibnamefont
  {Christensen}}, \bibinfo {author} {\bibfnamefont {M.}~\bibnamefont
  {Du{\l}ak}}, \bibinfo {author} {\bibfnamefont {J.}~\bibnamefont {Friis}},
  \bibinfo {author} {\bibfnamefont {M.~N.}\ \bibnamefont {Groves}}, \bibinfo
  {author} {\bibfnamefont {B.}~\bibnamefont {Hammer}}, \bibinfo {author}
  {\bibfnamefont {C.}~\bibnamefont {Hargus}}, \emph {et~al.},\ }\bibfield
  {title} {\bibinfo {title} {The atomic simulation environment—a python
  library for working with atoms},\ }\href@noop {} {\bibfield  {journal}
  {\bibinfo  {journal} {J. Condens. Matter Phys.}\ }\textbf
  {\bibinfo {volume} {29}},\ \bibinfo {pages} {273002} (\bibinfo {year}
  {2017})}\BibitemShut {NoStop}%
\bibitem [{\citenamefont {Mortensen}\ \emph {et~al.}(2024)\citenamefont
  {Mortensen}, \citenamefont {Larsen}, \citenamefont {Kuisma}, \citenamefont
  {Ivanov}, \citenamefont {Taghizadeh}, \citenamefont {Peterson}, \citenamefont
  {Haldar}, \citenamefont {Dohn}, \citenamefont {Sch{\"a}fer}, \citenamefont
  {J{\'o}nsson} \emph {et~al.}}]{Mortensen2024}%
  \BibitemOpen
  \bibfield  {author} {\bibinfo {author} {\bibfnamefont {J.~J.}\ \bibnamefont
  {Mortensen}}, \bibinfo {author} {\bibfnamefont {A.~H.}\ \bibnamefont
  {Larsen}}, \bibinfo {author} {\bibfnamefont {M.}~\bibnamefont {Kuisma}},
  \bibinfo {author} {\bibfnamefont {A.~V.}\ \bibnamefont {Ivanov}}, \bibinfo
  {author} {\bibfnamefont {A.}~\bibnamefont {Taghizadeh}}, \bibinfo {author}
  {\bibfnamefont {A.}~\bibnamefont {Peterson}}, \bibinfo {author}
  {\bibfnamefont {A.}~\bibnamefont {Haldar}}, \bibinfo {author} {\bibfnamefont
  {A.~O.}\ \bibnamefont {Dohn}}, \bibinfo {author} {\bibfnamefont
  {C.}~\bibnamefont {Sch{\"a}fer}}, \bibinfo {author} {\bibfnamefont
  {E.~{\"O}.}\ \bibnamefont {J{\'o}nsson}}, \emph {et~al.},\ }\bibfield
  {title} {\bibinfo {title} {GPAW: An open python package for electronic
  structure calculations},\ }\href@noop {} {\bibfield  {journal} {\bibinfo
  {journal} {J. Chem. Phys.}\ }\textbf {\bibinfo {volume}
  {160}},\ \bibinfo {pages} {092503}  (\bibinfo {year} {2024})}\BibitemShut {NoStop}%
\bibitem [{\citenamefont {Gjerding}\ \emph {et~al.}(2021)\citenamefont
  {Gjerding}, \citenamefont {Skovhus}, \citenamefont {Rasmussen}, \citenamefont
  {Bertoldo}, \citenamefont {Larsen}, \citenamefont {Mortensen},\ and\
  \citenamefont {Thygesen}}]{Gjerding2021}%
  \BibitemOpen
  \bibfield  {author} {\bibinfo {author} {\bibfnamefont {M.}~\bibnamefont
  {Gjerding}}, \bibinfo {author} {\bibfnamefont {T.}~\bibnamefont {Skovhus}},
  \bibinfo {author} {\bibfnamefont {A.}~\bibnamefont {Rasmussen}}, \bibinfo
  {author} {\bibfnamefont {F.}~\bibnamefont {Bertoldo}}, \bibinfo {author}
  {\bibfnamefont {A.~H.}\ \bibnamefont {Larsen}}, \bibinfo {author}
  {\bibfnamefont {J.~J.}\ \bibnamefont {Mortensen}},\ and\ \bibinfo {author}
  {\bibfnamefont {K.~S.}\ \bibnamefont {Thygesen}},\ }\bibfield  {title}
  {\bibinfo {title} {Atomic simulation recipes: A python framework and library
  for automated workflows},\ }\href@noop {} {\bibfield  {journal} {\bibinfo
  {journal} {Comput. Mater. Sci.}\ }\textbf {\bibinfo {volume}
  {199}},\ \bibinfo {pages} {110731} (\bibinfo {year} {2021})}\BibitemShut
  {NoStop}%
\end{thebibliography}

%

\end{document}